\documentclass[11pt,a4paper]{article}
\usepackage[T1]{fontenc}
\usepackage[utf8]{inputenc}
\usepackage{lmodern}
\usepackage{amsmath,amssymb,amsthm}
\usepackage{graphicx}
\usepackage{booktabs}
\usepackage[margin=2.4cm]{geometry}
\usepackage{microtype}

\newcommand{\E}{\mathbb{E}}
\newcommand{\Var}{\operatorname{Var}}

\newcommand{\Corr}{\operatorname{Corr}}
\newcommand{\Mult}{\operatorname{Mult}}
\newcommand{\sgn}{\operatorname{sgn}}
\newcommand{\Bin}{\operatorname{Bin}}
\newcommand{\erf}{\operatorname{erf}}
\newcommand{\KS}{\mathrm{KS}_{1/2}}
\newcommand{\AD}{\mathrm{AD}_{1/2}}
\newcommand{\LL}{\mathrm{L}^2_{1/2}}

\newtheorem{theorem}{Theorem}[section]

\newtheorem{proposition}[theorem]{Proposition}

\newtheorem*{propositionU}{Proposition}
\theoremstyle{definition}
\newtheorem{remark}[theorem]{Remark}
\newtheorem{observation}[theorem]{Numerical observation}

\title{The exact null distribution of the discrete total variation}
\author{
Nikola Bani\'c\\
Gideon Brothers, 10000 Zagreb, Croatia\\
\texttt{nbanic@gmail.com}
\and
Neven Elezovi\'c\\
Department of Applied Mathematics,\\
Faculty of Electrical Engineering and Computing,\\
University of Zagreb, 10000 Zagreb, Croatia\\
\texttt{neven.elezovic@fer.hr}
}
\date{}

\begin{document}
\maketitle

\begin{abstract}
\noindent
The discrete total variation (DTV) $d=\sum_i|X_{i+1}-X_i|$ of a multinomial histogram,
and its version $d_{\mathbf r}$ centered at a specified rational null $\mathbf r$, have
exact null laws that can be computed by dynamic programming. We describe the recurrence
for both nulls in one generic form, give a running-sum version that needs $O(nN^3)$
arithmetic operations for $d$ instead of $O(nN^4)$ (and $O(nLN^3)$ for $d_{\mathbf r}$),
and prove an error bound for its floating-point version. We then determine the lattice on
which the exact law lives: for $r_i=q_i/L$ ($q_i$ positive integers with $\gcd_iq_i=1$),
$d_{\mathbf r}=0$ is possible if and only if $L\mid N$; for $n\ge3$ the support lies on a
lattice of step $s/L$, with $s$ an explicit gcd that depends on $N$ modulo $L$, and this is
the exact step for all large $N$, though not always for small $N$; we also give the parity
law and the maximum. For three cells the uniform law has a permanent period-$3$ sawtooth,
which we describe by a local limit theorem, and in general the masses of the residue
classes of $L\,d_{\mathbf r}$ modulo $L$ converge, along each class of $N$ modulo $L$, to
sums of Gaussian orthant probabilities. The exact cumulants agree numerically with the limits of the companion paper. Finally we calibrate continuous approximations against exact tables.
With exact finite-sample moments, a gamma-based Cornish--Fisher quantile with a half-span
lattice shift gives the exact critical value in five illustrative uniform cases, but not
uniformly on a wider grid; under general nulls (on $22$ profiles) it is exact in only $39$--$53\%$ of $182$
tables (the normal-based Cornish--Fisher quantile is exact more often at the levels $1\%$ and $0.1\%$), and it can
fail completely when some expected counts are small. We therefore recommend
the exact law, or at least an exact check of the tail, whenever the level must be
controlled.
\end{abstract}



\section{Introduction}
	\label{sec:intro}

Let $X=(X_1,\ldots,X_n)\sim\Mult(N;\mathbf p)$ be the counts of a histogram of $N$
observations over $n$ ordered cells, and let $\mathbf r$ be the null law, so that
$\mathbf p=\mathbf r$ under the null. The linear and cyclic discrete total variation
(DTV) statistics are
\[
	d=\sum_{i=1}^{n-1}|X_{i+1}-X_i|,\qquad
	d^c=\sum_{i=1}^{n}|X_{i+1}-X_i|\quad(X_{n+1}=X_1),
\]
and the statistic centered at the null is
\begin{equation}\label{eq:dr}
	d_{\mathbf r}=\sum_{i=1}^{n-1}\bigl|X_{i+1}-X_i-N(r_{i+1}-r_i)\bigr| ,
\end{equation}
so that $d_{\mathbf u}=d$ for the uniform null $\mathbf u$. The cyclic statistic $d^c$ is
treated in~\cite{banic_elezovic_p5}; here we only record its definition. This paper is about the exact
null laws of $d$ and $d_{\mathbf r}$: how they are computed, on which lattice they
live, and how well continuous laws approximate them. Throughout $\mathbf p=\mathbf r$,
$\Delta_i=r_{i+1}-r_i$, $D_i=X_{i+1}-X_i$ and $\tilde D_i=D_i-N\Delta_i$; $\alpha$ is
always a significance level, and the moment-matched gamma law has shape $\nu$ and scale
$\theta$. Throughout, a general null means a rational null with positive cells,
$r_i=q_i/L$ with integers $q_i\ge1$, $\gcd_iq_i=1$ and $L=\sum_iq_i$; the exact integer
algorithm needs a rational null, and the lattice results of Section~\ref{sec:lattice} need
the representation in lowest terms. The null, the cells, their order and the choice of
statistic ($d$, $d_{\mathbf r}$ or $d^c$) are fixed before the data are seen. The laws
computed here are not the laws of a procedure in which $\mathbf r$ is estimated from the
same sample (for
$\hat{\mathbf r}=X/N$ one has $d_{\hat{\mathbf r}}=0$
identically, since every centered edge vanishes), or in which the cells are chosen or
ordered after inspection. The limit theorems are for fixed $n$, a fixed null with positive
cells and $N\to\infty$; sparse or high-dimensional regimes, in which some $Nr_i$ stay
small, are not covered (see the end of Section~\ref{sec:cumulants}, and
Section~\ref{sec:calibration} for what can then go wrong).

The statistic $d$ is the statistic of the comb test of uniformity, introduced and
illustrated in~\cite{banic_elezovic_ct}. There its exact null law was computed by a
dynamic programme of cost $O(nN^4)$ for every pair $2\le N,n\le500$ (all $249{,}001$
pairs), $29$ candidate families were
fitted to the exact null distribution functions by a discrete Cram\'er--von Mises criterion
and compared, and a gamma reference was recommended. The centered
statistic $d_{\mathbf r}$ for a non-uniform null, its exact null law by a weighted
dynamic programme, the sign locking of the raw statistic $d$ under a non-uniform null
and its resulting near-normal shape (when all $\Delta_i\ne0$) were introduced in~\cite{banic_elezovic_gct}.
(There, with integer weights $q_i$ and $r_i=q_i/L$, the integer statistic $S=L\,d_{\mathbf r}$ is
used, the slopes $\Delta_i$ are written $g_i$, and $d_{\mathbf u}$ denotes the
uncentered $d$.) The DTV was introduced by us as an outlier measure for histograms
in~\cite{banic_elezovic_tvor}. The algorithms and the tables were introduced
and illustrated in~\cite{banic_elezovic_ct,banic_elezovic_gct}; here we give the
mathematical development: the structure of the computation, its accuracy, the
lattice of the exact law, and the calibration of approximations against it.

\paragraph{Place in the series.}
The mean and the variance of $d$ and $d_{\mathbf r}$ are
in~\cite{banic_elezovic_p1,banic_elezovic_p2,banic_elezovic_p3,banic_elezovic_p4}, and
the cyclic statistic in~\cite{banic_elezovic_p5}. The companion
paper~\cite{banic_elezovic_p07} derives the limit law of $d_{\mathbf r}/\sqrt N$, its cumulants (the
third in closed form for every null $\mathbf r$) and the tail approximation \emph{analytically};
the present paper treats the same laws by \emph{exact computation}. The two overlap on
purpose, the companion paper giving the analytic basis and this paper the exact
computational procedure: each states the other's numbers as a check and cites it, and neither
re-derives the other. Complete reference tables of exact quantiles and the software will
be published with~\cite{banic_elezovic_p11}; here we print only selected values. Power against
alternatives (combs, ramps, tents) is the subject of~\cite{banic_elezovic_p10}.
\begin{center}\small
\begin{tabular}{lll}
\toprule
 & uniform null $\mathbf u$ & general null $\mathbf r$\\
\midrule
moments & \cite{banic_elezovic_p1,banic_elezovic_p2,banic_elezovic_p5} & \cite{banic_elezovic_p3,banic_elezovic_p4}\\
limit law, cumulants, tails (analytic) & \cite{banic_elezovic_p07} & \cite{banic_elezovic_p07}\\
algorithm, lattice, calibration (exact) & this paper & this paper\\
complete tables, software & \cite{banic_elezovic_p11} & \cite{banic_elezovic_p11}\\
\bottomrule
\end{tabular}
\end{center}

\paragraph{Relation to the literature.}
Exact null laws of multinomial goodness-of-fit statistics have been computed since Good,
Gover and Mitchell~\cite{ggm1970}, for $X^2$ and the likelihood ratio in the equiprobable
case. For decomposable statistics, sums of functions of the single cells, the law can be
obtained by inverting a characteristic function with the fast Fourier transform (Baglivo,
Olivier and Pagano~\cite{baglivo1992}; for the likelihood-ratio statistic Keich and
Nagarajan~\cite{keich2006} stabilize it by exponential shifts) or by generating functions built cell after cell
(Selivanov~\cite{selivanov2006}); Hirji~\cite{hirji1997comparison} compares such
algorithms, Filina and Zubkov~\cite{filina2008exact} compute the law of Pearson's $X^2$
by a Markov chain (for uniform histograms see also our~\cite{zds}), and Zubkov and Filina~\cite{zubkov2020computation} give a general
Markov-chain scheme for statistics updated from a carried state, close in spirit to the
finite Markov-chain imbedding of Fu and Koutras~\cite{fu1994distribution}. The network
algorithm of Mehta and Patel~\cite{mehta1983network} serves conditional inference in
contingency tables, and Resin~\cite{resin2023} computes exact multinomial $p$-values by
enumerating an acceptance region instead of the whole law. Bonetti, Cirillo and
Ogay~\cite{bonetti2019} compute, under the equiprobable null, exact laws of functionals
of the counts that are not decomposable but are invariant under permutations of the
cells (maximum, minimum, range). For ordered cells, exact null laws of the
Kolmogorov--Smirnov statistic for grouped or discrete data are computed
in~\cite{pettitt_stephens1977,dimitrova2020}; that statistic depends on the cumulative
counts, not on differences of adjacent counts. The DTV is neither decomposable nor permutation
invariant: it couples adjacent cells, so the product of one-cell transforms used for
decomposable statistics does not apply. A cell-by-cell computation, of the law, of its
generating function or of its characteristic function, must carry the count of the
previous cell (or an equivalent separator state); a transform of the joint law of the
$n-1$ edges (Section~\ref{sec:algorithm}) is another exact route. Exact recursions through separator states
are also known for ordered window statistics: Kuriki, Takahashi and
Hara~\cite{kuriki2018} compute exact $p$-values of temporal and spatial scan statistics
under the multinomial law by a recursion based on the Markov property of a chordal graph;
their window events differ from the overlapping adjacent absolute differences of the DTV,
whose accumulated sum needs the running sums below to be computed efficiently. To our
knowledge, the exact null law of $\sum_i|X_{i+1}-X_i|$ under a one-way multinomial law
was first published in~\cite{banic_elezovic_ct} (uniform null), and its extension to a
rational null is in the preprint~\cite{banic_elezovic_gct}; for related statistics of
normal samples, such as the mean successive difference, the literature studies moments
and limits, not this finite multinomial law. What is new here is therefore not exact
multinomial computation, recursion through a separator, or transform inversion as such,
but their adaptation to this nearest-neighbour statistic: the running sums (a two-pass
recursion for the two-sided kernel $u^{|A(b-a)-c|}$, as in one-dimensional distance
transforms~\cite{felzenszwalb2012}), a floating-point bound for them (by the standard
forward error analysis of sums and products of non-negative
numbers~\cite{higham2002}), and the lattice and residue structure of the law.
Fitting a scaled $\chi^2$ (gamma) law by two moments to a quadratic-form or
variance-type statistic goes back to Welch~\cite{welch1938},
Satterthwaite~\cite{satterthwaite1946}, Patnaik~\cite{patnaik1949} and
Box~\cite{box1954}; for the bounded statistic of Durbin and Watson a beta law fitted by
the first two moments was proposed in~\cite{durbin_watson1950,durbin_watson1951}.
Under the uniform null the limit of
$d/\sqrt N$ is, up to a factor, the mean successive difference of a normal sample
(Hoel~\cite{hoel1946}, Kamat~\cite{kamat1953}; see~\cite[Theorem~3.1]{banic_elezovic_p07}).
For the tail we use the Cornish--Fisher quantile~\cite{cornish_fisher1938} in the
gamma-based form of~\cite[eq.~(16)]{banic_elezovic_p07}, a heuristic gamma-anchored
correction rather than an expansion of Cornish--Fisher type about the gamma law in the
sense of Hill and Davis~\cite{hill_davis1968},
and as a tail-weighted distance an Anderson--Darling-type
discrepancy (cf.~\cite{anderson_darling1952}).

\paragraph{Contributions.}
\begin{enumerate}
\item One generic recurrence for $d$, for $d$ under a weighted null and for
$d_{\mathbf r}$ at a rational null (Section~\ref{sec:algorithm}); a running-sum form with
$O(nN^2V_{\max})$ arithmetic operations, that is $O(nN^3)$ for $d$ and $O(nLN^3)$ for
$d_{\mathbf r}$ (Proposition~\ref{prop:runsum}); and a floating-point version with a
proved error bound (Proposition~\ref{prop:float}): when no underflow occurs, a relative error bound of order $nN$ units of roundoff for every positive probability and every positive
tail of $d$, when the binomial weights are correctly rounded, while probabilities that are
exactly zero are computed as zero; with underflow, a mixed bound, relative plus an
absolute term; for tails of $d_{\mathbf r}$ the
bound grows with the number of values summed (in IEEE arithmetic, with the number of
support points summed).
\item The lattice of $d_{\mathbf r}$ for $n\ge3$ (Theorem~\ref{thm:lattice}): $d_{\mathbf r}=0$ with positive probability if and only if $L\mid N$, a containing lattice of step $s/L$ that is exact for $N\ge N_0$ but not
for every $N$, the parity law and the maximum. Under the uniform null the
support and parity are those of~\cite[Proposition~2.1]{banic_elezovic_p07}; we add the point mass at $0$, the masses next to the holes and the size of the parity
deviation.
\item The local limit law for $n=3$, a permanent period-$3$ sawtooth
(Theorem~\ref{thm:n3local}), and the limit of the residue classes of $L\,d_{\mathbf r}$
modulo $L$ as Gaussian orthant probabilities (Theorem~\ref{thm:residues}).
\item The exact cumulants of $d_{\mathbf r}$ against the analytic limits
of~\cite{banic_elezovic_p07} (Section~\ref{sec:cumulants}).
\item A calibration of continuous approximations on exact tables with
continuity-corrected measures (Section~\ref{sec:calibration}): the half-span rule with the
gamma-based Cornish--Fisher quantile is exact in $15/15$ illustrative uniform cells
(pairs of case and level) with
$n=10,50$ (but not at small $n$, \cite[Section~7.4]{banic_elezovic_p07}); under the uniform null
the moment-matched beta on $[0,2N]$ and the gamma law have the same limit as $N\to\infty$ for fixed $n$
(Proposition~\ref{prop:beta-gamma}), and a beta whose support is fitted to the exact
skewness is exact in $15/15$; under a general null the $\Gamma$+CF half-span rule (with
exact moments) is exact in only $39$--$53\%$ of $182$ tables and can fail completely for
profiles with small expected counts; the normal Cornish--Fisher quantile with the same rule
ties with it at $5\%$ and is better at $1\%$ and $0.1\%$ (at $0.1\%$ exact in $102$ against $71$ tables,
largest size $1.12\alpha$ against $1.49\alpha$), but neither controls the size, so the exact law is recommended there, and under
any null when the size must not exceed $\alpha$.
\end{enumerate}
Statements in theorem environments are proved. Statements labelled \emph{numerical
observation} are read from exact tables and are not proved.

\paragraph{Organisation.}
Section~\ref{sec:algorithm} gives the algorithm, Section~\ref{sec:lattice} the lattice
structure, Section~\ref{sec:cumulants} the exact cumulants against the analytic limits,
Section~\ref{sec:calibration} the calibration, and Section~\ref{sec:open} open problems,
followed by a short conclusion.

\section{The exact algorithm}
	\label{sec:algorithm}

\subsection{One recurrence for both nulls}

Let $q_1,\ldots,q_n$ be positive integers with $\gcd_iq_i=1$, $L=\sum_iq_i$ and
$r_i=q_i/L$; every rational null with positive cells has this form, and the uniform null
is $q_i=1$, $L=n$. (A null with zero cells can be run through the exact recurrence below
with $q_i=0$ and the convention $0^0=1$; in its floating-point version below, the
conditional weight is then $\mathbf 1\{b=0\}$ when $q_j+\cdots+q_n=0$, and this rule takes
precedence over every other one, including the last-cell rule $\mathbf 1\{b=N-\ell\}$. The
output then collects only states with total count $N$, so mass that additive roundoff may
put on impossible states (total below $N$ with no positive cell left) is dropped; this
convention is the one covered by Proposition~\ref{prop:float}. A zero cell is then
a cell with $X_i=0$ almost surely, which is not the same as deleting it.
Theorem~\ref{thm:lattice}, the limit theorems of
Section~\ref{sec:lattice}, the limits of Section~\ref{sec:cumulants} and the calibration of Section~\ref{sec:calibration} assume
positive cells. Irrational nulls are outside the exact integer algorithm, although the
floating-point version below accepts real $r_i$ for the raw statistic $d$, with the
weights $\Pr\bigl(\Bin(N-\ell,\,r_j/(r_j+\cdots+r_n))=b\bigr)$.)
Write $\delta_i=q_{i+1}-q_i$, so that $\Delta_i=\delta_i/L$. The statistics $d$ (under any null) and $L\,d_{\mathbf r}$ are of the form
\begin{equation}\label{eq:generic}
	V=\sum_{i=1}^{n-1}\bigl|A(X_{i+1}-X_i)-c_i\bigr| ,
\end{equation}
with integers $A\ge1$ and $c_i$: $A=1$, $c_i=0$ gives $V=d$ (for any null), and $A=L$,
$c_i=N\delta_i$ gives
\[
	V=L\,d_{\mathbf r}=\sum_{i=1}^{n-1}\bigl|L(X_{i+1}-X_i)-N\delta_i\bigr|\in\mathbb Z_{\ge0}.
\]
For $0\le a\le t\le N$ and $1\le j\le n$ let $C_j(t,a,v)$ be the total weight of the
placements $x_1,\ldots,x_j\ge0$ with $x_1+\cdots+x_j=t$, $x_j=a$ and
$\sum_{i<j}|A(x_{i+1}-x_i)-c_i|=v$, where a placement has weight
$\frac{N!}{x_1!\cdots x_j!\,(N-t)!}\prod_{i\le j}q_i^{x_i}$. Then
$C_1(t,t,0)=\binom Ntq_1^t$, and for $j\ge2$
\begin{equation}\label{eq:dp}
	C_j(t,b,v)=\binom{N-t+b}{b}q_j^{\,b}
	\sum_{a=0}^{t-b}C_{j-1}\bigl(t-b,a,v-|A(b-a)-c_{j-1}|\bigr),
\end{equation}
with $C_{j-1}=0$ outside its range, and
\[
	\Pr(V=v)=L^{-N}\sum_{a=0}^NC_n(N,a,v).
\]
This is the recurrence of~\cite[eq.~(4)]{banic_elezovic_ct} for the uniform null and
of~\cite[eq.~(4)]{banic_elezovic_gct} for $d_{\mathbf r}$; its correctness is the
multinomial identity $\binom N{x_1,\ldots,x_n}=\prod_j\binom{N-x_1-\cdots-x_{j-1}}{x_j}$
and $\sum_x\binom N{x}\prod_iq_i^{x_i}=L^N$. Equivalently, the probability generating
function of $V$ is a product of transfer matrices along the path of cells, with kernels
$u^{|A(b-a)-c_i|}$; the count of the last cell is the separator. The output is the whole
law; tails and critical values are read off it. As in the count of~\cite{banic_elezovic_ct} for $d$, there are $O(N^2V_{\max})$ states per cell, each a sum of $O(N)$ terms, where
$V_{\max}\le2AN+\sum_i|c_i|$ bounds $V$; for $d$ this is $O(nN^4)$ arithmetic operations
and, keeping two layers, $O(N^3)$ memory.

\begin{remark}\label{rem:n3}
For $n=3$ no recurrence is needed:
$d_{\mathbf r}=\max\bigl(|X_3-X_1-N(r_3-r_1)|,\,3|X_2-Nr_2|\bigr)$
(\cite[Proposition~4.14(a)]{banic_elezovic_p07}; directly, $|a|+|b|=\max(|a+b|,|a-b|)$ for
the two centered edges $a,b$, with $a+b=X_3-X_1-N(r_3-r_1)$ and $a-b=3(X_2-Nr_2)$), and given $X_2=j$ the
count $X_1$ is $\Bin(N-j,r_1/(r_1+r_3))$, so the exact law is a double sum of $O(N^2)$
binomial terms. We use it for $n=3$ and $N$ up to $2400$ below.
\end{remark}

\subsection{Running sums}

Fix $j\ge2$ and $t-b=\ell$, and write $F_a(u)=\sum_vC_{j-1}(\ell,a,v)u^v$ for the generating
polynomial of a state, $c=c_{j-1}$ and $\tau=\lceil c/A\rceil$. By~\eqref{eq:dp} the
generating polynomial of $C_j(\ell+b,b,\cdot)$ is $\binom{N-\ell}{b}q_j^{\,b}H_b(u)$ with
\[
	H_b(u)=\sum_{a=0}^{\ell}u^{|A(b-a)-c|}F_a(u),\qquad 0\le b\le N-\ell .
\]
Computed term by term this costs $O(N)$ polynomial additions for each $b$.

\begin{proposition}[Running sums]\label{prop:runsum}
Put $F_a=0$ for $a\notin\{0,\ldots,\ell\}$. Then $H_b=P_b+Q_b$, where $P_b$ and $Q_b$ are
determined by $P_b=0$ for $b<\tau$, $Q_b=0$ for $b>\ell+\tau-1$, and
\[
	P_b=u^AP_{b-1}+u^{A\tau-c}F_{b-\tau},\qquad
	Q_b=u^AQ_{b+1}+u^{c-A(\tau-1)}F_{b-\tau+1}.
\]
Both exponents are non-negative. The recursion for $P$ starts at $b=\tau$, which may be
negative, and the recursion for $Q$ starts at $b=\ell+\tau-1$, which may exceed $N-\ell$;
both are run through these auxiliary values of $b$. Hence, if $|c_i|\le AN$ for all $i$
(as for $d$, and for $L\,d_{\mathbf r}$ since $|\delta_i|\le L$, strictly for positive
cells), all $H_b$, $0\le b\le N-\ell$,
are obtained with $O(N)$ shifts and additions of polynomials of degree at most $V_{\max}$, the whole
law with $O(nN^2V_{\max})$ arithmetic operations, that is $O(nN^3)$ for $d$ and $O(nLN^3)$ for
$d_{\mathbf r}$, and with memory $O(N^2V_{\max})$.
\end{proposition}

\begin{proof}
The exponent $A(b-a)-c$ is non-negative if and only if $a\le b-\tau$. Split
$H_b=P_b+Q_b$ with $P_b=\sum_{a\le b-\tau}u^{A(b-a)-c}F_a$ and
$Q_b=\sum_{a\ge b-\tau+1}u^{c-A(b-a)}F_a$. Then
\[
	u^AP_{b-1}=\sum_{a\le b-1-\tau}u^{A(b-a)-c}F_a,\qquad
	u^AQ_{b+1}=\sum_{a\ge b-\tau+2}u^{c-A(b-a)}F_a,
\]
which are $P_b$ without its term $a=b-\tau$ and $Q_b$ without its term $a=b-\tau+1$. Finally $A\tau-c\ge0$ and $c-A(\tau-1)>0$ by the definition of
$\tau$. The cost count follows: there are $O(nN)$ pairs $(j,\ell)$, each needing
$O(N+|\tau|)=O(N)$ polynomial operations of length $O(V_{\max})$, since $|\tau|\le N$
when $|c|\le AN$.
\end{proof}

\paragraph{Implementation.}
In pseudocode, the law is computed as follows; we checked this procedure against
brute-force enumeration of all compositions, for raw and centered statistics and for
negative and positive $\tau$ ($118$ cases, $12$ profiles, $N\le7$; this check of the
pseudocode as written is separate from the three-way verification of Section~\ref{sec:verification}). Write
$F^{(j)}_{t,a}(u)=\sum_vC_j(t,a,v)\,u^v$.
\begin{enumerate}
\item Put $F^{(1)}_{t,t}=\binom Ntq_1^{\,t}$ for $0\le t\le N$.
\item For $j=2,\ldots,n$ and each $\ell=0,\ldots,N$, with $c=c_{j-1}$,
$\tau=\lceil c/A\rceil$ and $F_a=F^{(j-1)}_{\ell,a}$ ($F_a=0$ for $a\notin\{0,\ldots,\ell\}$):
\begin{enumerate}
\item put $P=0$ and, for $b=\tau,\tau+1,\ldots,N-\ell$, set
$P\leftarrow u^AP+u^{A\tau-c}F_{b-\tau}$, storing $P_b=P$ when $b\ge0$;
\item put $Q=0$ and, for $b=\ell+\tau-1,\ell+\tau-2,\ldots,0$, set
$Q\leftarrow u^AQ+u^{c-A(\tau-1)}F_{b-\tau+1}$, storing $Q_b=Q$ when $b\le N-\ell$
(unstored $P_b$, $Q_b$ are $0$);
\item for $0\le b\le N-\ell$ (only $b=N-\ell$ when $j=n$) put
$F^{(j)}_{\ell+b,b}=\binom{N-\ell}{b}q_j^{\,b}(P_b+Q_b)$.
\end{enumerate}
Then discard layer $j-1$.
\item Output $\Pr(V=v)=L^{-N}[u^v]\sum_aF^{(n)}_{N,a}(u)$ for all $v$.
\end{enumerate}
The polynomials can be stored as arrays indexed by the exponent of $u$; for
$d_{\mathbf r}$ only the occupied residue classes modulo $L$ need to be stored (see the
next paragraph). The variant for a single tail is described at the end of
Section~\ref{sec:calibration}.

For $d_{\mathbf r}$ the factor $L$ can be reduced: since $y_i=LD_i-N\delta_i\equiv-N\delta_i
\pmod L$, after $j-1$ edges the partial sum $\sum_{i<j}|y_i|$ lies in at most
$\min(L,2^{j-1})$ residue classes modulo $L$~\cite{banic_elezovic_gct}, and the final law
in one coset of $s\mathbb Z$ (Theorem~\ref{thm:lattice}(c) below). With sparse storage of
the occupied classes the cost becomes $O(n\,\kappa\,N^3)$ with $\kappa\le\min(L,2^{n-1})$.
As anecdotal context only, not a benchmark: in earlier single runs of our Python
implementation with exact big integers on a laptop, the uniform law at $(N,n)=(200,10)$
took $33$\,s with the running sums against $360$\,s with~\eqref{eq:dp}, and $2.6$\,s in the
floating-point version below, and the two-point slopes of $\log$(time) against $\log N$
between $N=80$ and $N=200$ were $4.0$ and $5.0$; the hardware and operating system of these
runs, their variability and their peak memory were not recorded, and the runs were not
repeated. (The exact and floating-point checks of this paper were later rerun
with CPython~3.12.10, NumPy~2.2.6 and SciPy~1.15.3 on a laptop with an Intel Core
i7-12700T and $32$\,GB RAM; the timings above do not come from these reruns.) The counts above are counts of
arithmetic operations; with exact integers the coefficients have $O(N\log L)$ bits, which
accounts for the extra factor $N$ in the observed times. The direct and the running-sum
exact-integer implementations give identical integer tables.

Other exact routes exist. One observation in cell $j$ moves the edge vector
$D=(D_1,\ldots,D_{n-1})$ by $-\mathbf e_1$ ($j=1$), by $\mathbf e_{j-1}-\mathbf e_j$
($2\le j\le n-1$) or by $\mathbf e_{n-1}$ ($j=n$), so $L^N\Pr(D=\mathbf d)$ is the
coefficient of $\mathbf z^{\mathbf d}$ in
$\bigl(q_1z_1^{-1}+\sum_{j=2}^{n-1}q_jz_{j-1}z_j^{-1}+q_nz_{n-1}\bigr)^N$. A
multidimensional discrete Fourier transform of length $2N+1$ per coordinate recovers the
joint law of $D$ (exactly by finite-field fast Fourier transforms~\cite{pollard1971}
modulo primes that have roots of unity of a suitable order, the integer coefficients being
recovered by the Chinese remainder theorem once the product of the moduli exceeds $L^N$; in
floating point up to rounding), and hence that of $d_{\mathbf r}$, with
$O(n(2N+1)^{n-1}\log N)$ ring operations per modulus: there is no factor $L$ in this
count (the coefficients still have $O(N\log L)$ bits), but the cost is exponential in $n$. Since $D$ determines $X$ when $N$ is
fixed, the joint law of $D$ is the multinomial law itself on $\binom{N+n-1}{n-1}\le(N+1)^{n-1}$ points, so this route is
never cheaper than enumeration of the compositions; its only merit is that its operation count has no factor $L$. When only one tail is needed, the acceptance region
$\{d_{\mathbf r}<c\}$ consists of the integer points of the convex polytope
$\{x\ge0,\ \sum_ix_i=N,\ d_{\mathbf r}(x)\le c'\}$, $c'$ the largest support point below $c$ ($d_{\mathbf r}$ is convex
in $x$), and it can be enumerated directly. The algorithm of Resin~\cite{resin2023}, which grows the acceptance region
around the expected counts, is exact for weakly quasi M-convex statistics; for $d$ this property fails when $n\ge4$:
at $(N,n)=(6,4)$ the set $\{d\le1\}=\{(1,1,2,2),(2,2,1,1)\}$ is not connected by moves of one observation.

The direct recurrence~\eqref{eq:dp} can also be run with sparse storage, each state
$C_j(t,a,\cdot)$ being kept as the list of its non-zero coefficients, keyed by $v$; its cost is then
controlled by the size of the final support instead of the range $V_{\max}$, as follows.

\begin{propositionU}[Output-sensitive exact law]
Let $q_i\ge1$ for all $i$, let $V$ be as in~\eqref{eq:generic} and let $m$ be the number of
support points of $V$. In~\eqref{eq:dp} with sparse storage, computed for $j=n$ only at
$t=N$, every state has at most $m$ non-zero coefficients. Hence the whole law is obtained
with $O(nN^3m)$ additions and multiplications of integers of $O(N\log L)$ bits, together
with $O(nN^3m\log m)$ comparisons of keys of $O(\log(nLN))$ bits in balanced search trees,
and with $O(N^2m)$ stored coefficients per layer. The algorithm does not need $m$ as an
input, and its bit cost is polynomial in $n$, $N$, $\log L$ and $m$.
\end{propositionU}

\begin{proof}
Let $j\le n-1$ and let $v$ be a key of the state $(j,t,a)$, that is, $C_j(t,a,v)>0$, so
that some placement $x_1,\ldots,x_j$ with $x_1+\cdots+x_j=t$ and $x_j=a$ has partial cost
$v$ (all coefficients are sums of positive terms, so none cancels). Complete it by
$x_{j+1}=\cdots=x_{n-1}=0$ and $x_n=N-t$. Since all $q_i\ge1$, this composition of $N$ has
positive probability, and its value of $V$ is $v+C_{j,t,a}$, where
$C_{j,t,a}=\sum_{i=j}^{n-1}|A(x_{i+1}-x_i)-c_i|$ depends only on $j$, $t$ and $a$. The
map $v\mapsto v+C_{j,t,a}$ is injective and takes the keys of the state into the support
of $V$, so there are at most $m$ keys. For $j=n$ and $t=N$ the keys are support points.
There are $O(nN^2)$ states, each a sum over $O(N)$ predecessor states with at most $m$ keys
each, and the target state has at most $m$ keys at every stage of the summation, since
every key inserted is a key of the completed state. Every coefficient is at most $L^N$,
and every key at most $V_{\max}\le2AN+\sum_i|c_i|=O(nLN)$.
\end{proof}

Since $m\le V_{\max}+1$, up to the factor $\log m$ this is never worse than the dense
direct recurrence, and it is cheaper than the running sums when $m$ is small compared with
$V_{\max}/N$, for instance when $L$ is large compared with the number
$\binom{N+n-1}{n-1}\ge m$ of compositions.

\subsection{Floating point}

The recurrence can be run in floating point with probabilities instead of counts:
replace the initial weight $\binom Ntq_1^{\,t}$ by $\Pr(\Bin(N,q_1/L)=t)$ and the weight
$\binom{N-\ell}{b}q_j^{\,b}$ by the conditional binomial probability
$\Pr\bigl(\Bin(N-\ell,\,q_j/(q_j+\cdots+q_n))=b\bigr)$ (and by $\mathbf 1\{b=N-\ell\}$ for
$j=n$). Then each state carries the joint probability of its partial configuration, and
the output is $\Pr(V=v)$ directly.

\begin{proposition}[Accuracy of the floating-point version]\label{prop:float}
Let $|c_i|\le AN$ for all $i$, let $\tau^*=\max_j|\lceil c_j/A\rceil|$ ($\tau^*=0$ for $d$),
\[
	K=(n-1)(N+\tau^*+3)+N+V_{\max}+1,\qquad \rho=(1+\epsilon_{\mathrm m})^K(1+\eta)^n ,
\]
and let $M$ and $W$ be the numbers of floating-point multiplications and of binomial weights
used (so $M\le(n-1)\binom{N+2}2(V_{\max}+1)$ and $(n-1)(N+1)\le W\le n\binom{N+2}2$).
Suppose that every floating-point operation on non-negative numbers returns a non-negative
number, that an addition returns $(x+y)(1+\delta)$ and a multiplication $xy(1+\delta)+e$,
with $|\delta|\le\epsilon_{\mathrm m}$ and $|e|\le\omega$, that every binomial weight $w$ is
computed as a non-negative number $w(1+\delta')+e'$ with $|\delta'|\le\eta$ and $|e'|\le\omega$, where
$0\le\epsilon_{\mathrm m},\eta<1$, and that $\rho\le2$ and $(M+W)\,\omega\le\frac1{100}$.
Then for every set
$S$ of values of $V$ (a single value or a tail, summed directly over its values, not as one minus a computed
distribution function)
\[
	\bigl|\hat P(V\in S)-\Pr(V\in S)\bigr|\le(\rho-1)\Pr(V\in S)+5(M+W)\,\omega .
\]
In particular, if no underflow occurs ($\omega=0$), then
$|\hat P(V\in S)/\Pr(V\in S)-1|\le\rho-1$ whenever $\Pr(V\in S)>0$, and $\hat P(V=v)=0$
whenever $\Pr(V=v)=0$.
\end{proposition}

\begin{proof}
Let first $\omega=0$. All inputs are non-negative, and the algorithm forms only sums and products, without
subtraction. By induction over the operations, every computed quantity is a sum of
products $\prod_kw_k$ of exact non-negative weights, each product multiplied by a factor
in $[(1-\epsilon_{\mathrm m})^{K'}(1-\eta)^{n'},(1+\epsilon_{\mathrm m})^{K'}(1+\eta)^{n'}]$, where $K'$ is the number of
floating-point operations and $n'$ the number of binomial weights on the path of that
product. Since all products are non-negative, the computed sum lies between the exact
sum multiplied by the smallest and by the largest of these factors. Along any path, one
cell contributes one binomial weight, one multiplication, at most $N+\tau^*+1$ additions
in the running sums of Proposition~\ref{prop:runsum} (which start at $b=\tau$ or
$b=\ell+\tau-1$) and one addition $P_b+Q_b$; the final sum over $a$ and a tail sum add at
most $N$ and $V_{\max}+1$ more. A value $v$ with $\Pr(V=v)=0$ is reached by no path, and
shifts and sums never create one, so $\hat P(V=v)=0$. This gives the bound with the
stated $K$; the lower deviation is dominated by the upper one, since
$1-\prod_k(1-x_k)\le\prod_k(1+x_k)-1$ for $x_k\ge0$.

For $\omega>0$ compare the computation with two exact computations in which every
operation takes its worst case: an upper one, in which $x+y$ becomes
$(x+y)(1+\epsilon_{\mathrm m})$, $xy$ becomes $xy(1+\epsilon_{\mathrm m})+\omega$ and $w$
becomes $w(1+\eta)+\omega$ (the last-cell weights $\mathbf 1\{b=N-\ell\}$ and the
zero-tail weights $\mathbf 1\{b=0\}$ are exact), and a
lower one, in which they become
$(x+y)(1-\epsilon_{\mathrm m})$, $\max\{0,xy(1-\epsilon_{\mathrm m})-\omega\}$ and
$\max\{0,w(1-\eta)-\omega\}$. These maps are non-decreasing in their arguments, so by
induction every computed value lies between its lower and its upper counterpart. In the
upper computation one cell multiplies the total mass by at most
$(1+\epsilon_{\mathrm m})^{k_j}\bigl((1+\eta)+(N+1)\omega\bigr)$, where $k_j$ is the number
of floating-point operations on a path through the cell, since the exact weights leaving a
state sum to $1$ over at most $N+1$ values of $b$, the weights of the last cell are exact,
and the running sums put each coefficient into each $H_b$ once. Hence the product of these
factors is at most $G=\rho\,e^{(n-1)(N+1)\omega}\le2e^{1/100}$. Write the upper computation
as its version with $\omega=0$ plus the terms created by $\omega$. A telescoping sum over
the cells shows that these terms have total mass at most
$G\bigl(\rho(n-1)(N+1)+M\bigr)\omega\le5(M+W)\omega$, because $(n-1)(N+1)\le W$. In the lower
computation every cell has mass gain at most $1$, and $\max\{0,\cdot\}$ is non-decreasing
and $1$-Lipschitz. Its deficit against its version with $\omega=0$ is therefore at most
the mass removed, $\bigl(M+(n-1)(N+1)\bigr)\omega\le(M+W)\omega$. Together with the case
$\omega=0$, applied to the two versions with $\omega=0$, this gives the bound.
\end{proof}

Some smallness condition on $(M+W)\omega$ is necessary: with $N=1$, $n=20$, exact
arithmetic and every weight returned as $2$ (so that $\omega=2$), the computed total mass
is $20\cdot2^{19}$ instead of $1$.

In IEEE double precision with rounding to nearest and gradual underflow, additions of
non-negative numbers are exact when the result is subnormal, so they satisfy the model
without an absolute term, with $\epsilon_{\mathrm m}=2^{-53}$, the unit roundoff (half the machine epsilon $2^{-52}$), and $\omega=2^{-1075}$. If
every weight is the correctly rounded quotient of the exact integers
$\binom mb q_j^{\,b}(T-q_j)^{m-b}$ and $T^m$, $T=q_j+\cdots+q_n$ (in CPython, which we used,
the quotient of two integers is correctly rounded), then
$\eta=\epsilon_{\mathrm m}$, and for the uniform null ($\tau^*=0$, $V_{\max}=2N$)
$\rho-1\approx(K+n)2^{-53}$ is $2.7\cdot10^{-13}$ at $(N,n)=(200,10)$ and $2.8\cdot10^{-11}$
at $N=n=500$; it grows like $nN$ units of roundoff. Underflow does occur for large $N$, for
instance $\Pr(X_1=N)=n^{-N}$ at $(500,10)$, but the absolute term $5(M+W)\omega$ is below
$10^{-312}$ at $N=n=500$, and below $10^{-296}$ even if all weights below $2^{-1022}$ are
set to $0$ (these two numbers depend only on $n$ and $N$ for the raw statistic $d$, under
any null, and the hypothesis $(M+W)\omega\le\frac1{100}$ holds there by about $300$ orders
of magnitude); so it matters only for probabilities far below any $p$-value of interest.
For a single probability the term $V_{\max}+1$ of $K$, which comes from the tail sum,
drops out, so $K=O(nN)$ for $d$ and $d_{\mathbf r}$ alike ($\tau^*\le N$). For a tail of
$d_{\mathbf r}$ summed over all $V_{\max}+1$ values it is $O(LN)$: for
$\mathbf q=(1,10000,1)$ and $N=500$, $V_{\max}=2.0\cdot10^7$ and
$\rho-1\approx2.2\cdot10^{-9}$. In IEEE arithmetic the addition of an exact zero is exact,
so a tail summed over its support points only (here $501$ of them) has a relative error of
the order of $(K-V_{\max}+|S\cap\operatorname{supp}V|)\,2^{-53}$, which is
$3\cdot10^{-13}$ in this example. The general-purpose
binomial probability of SciPy, which we used, has a larger error: on the weights of our
runs we measured $\eta$ up to $4\cdot10^{-13}$. Even so, the observed errors are small: against
the exact integer tables, the largest relative error of the probability mass function was
$1.4\cdot10^{-14}$ at $(N,n)=(100,10)$ and $9.4\cdot10^{-15}$ at $(200,10)$, down to
probabilities of $8\cdot10^{-40}$ and $1.3\cdot10^{-79}$, and the same for the upper tails;
with correctly rounded weights it is $7.9\cdot10^{-16}$ at $(100,10)$. In a run with $22$
underflowing weights ($n=3$, $\mathbf q=(1,1,8)$, raw $d$, $N=330$), every probability and
every upper tail above $10^{-300}$ had relative error below $6\cdot10^{-16}$ with correctly
rounded weights ($5\cdot10^{-14}$ with SciPy). So exact integers are needed only for certification, and even then only near the level: with
$B=5(M+W)\omega$ and $\rho<2$, and with correctly rounded weights (so that $\eta=\epsilon_{\mathrm m}$ is proved;
for the SciPy weights $\eta$ is only measured, and the bound below is then not a certificate), Proposition~\ref{prop:float} gives
$\max\{0,(\hat P-B)/\rho\}\le\Pr(V\in S)\le\min\{1,(\hat P+B)/(2-\rho)\}$, which decides whether a tail is at most
$\alpha$ unless $\hat P$ is within about $(\rho-1)\alpha+B$ of $\alpha$. The practical limits are memory and time: a layer holds (as array payload;
peak memory was not measured) $O(N^2V_{\max})=O(N^3)$ numbers, about $1$\,GB of double-precision numbers per layer at
$N=500$ (with triangular storage) and about $8$\,GB at $N=1000$, two layers being kept,
while the time is $O(nN^3)$ (an extrapolation of this count from $2.6$\,s at $(200,10)$
gives about half an hour at $N=n=500$; not measured). Within the range where the recurrence
fits in memory and in the available time, Monte Carlo is not needed. Outside it, the
Monte Carlo test~\cite{barnard1963,besag_clifford1989,dufour2006}, which compares the observed
value $t$ with $B$ independent samples $T_1,\ldots,T_B$ of the statistic under $\Mult(N;\mathbf r)$ and rejects
when $(1+\#\{b:T_b\ge t\})/(B+1)\le\alpha$, has size at most $\alpha$ for every $B$, ties included
(randomized tie-breaking gives size exactly $\alpha$ when $\alpha(B+1)$ is an integer), at a cost of $O(n+N)$ operations per sample.

\subsection{Verification}
	\label{sec:verification}

We implemented brute-force enumeration of all compositions, the
recurrence~\eqref{eq:dp} and the running sums of Proposition~\ref{prop:runsum}
independently of the original
programs of~\cite{banic_elezovic_ct,banic_elezovic_gct}. The three agree exactly on $196$ cases
($2\le n\le5$, $N\le8$, and $N\le6$ for $n=5$; the $13$ profiles $(1,1)$, $(1,1,1)$,
$(1,2,1)$, $(1,3,1)$, $(1,2,3)$, $(2,1,2)$, $(1,1,1,1)$, $(1,2,1,2)$, $(1,2,3,4)$,
$(3,1,2,1)$, $(1,1,1,1,1)$, $(1,2,3,4,5)$, $(1,3,2,4,1)$; raw and centered; a case is a
triple of profile, $N$ and statistic, so $196=16+80+64+36$ for $n=2,3,4,5$, and since $d_{\mathbf u}=d$ the
$30$ centered cases of the four uniform profiles repeat raw ones, leaving $166$ distinct laws), and every
total equals $L^N$. The running-sum version reproduces exactly (identical integers) the
exact uniform
tables at $(N,n)=(100,10)$ and $(200,10)$ used in~\cite{banic_elezovic_p07}, and the
exact tables of $d_{\mathbf r}$ for $\mathbf q=(1,2,1)$, $N=100$; $(1,2,3)$, $N=60$;
$(1,3,2,4,1)$, $N=50$; and $(1,2,3,4,5)$, $N=40$.

\section{Lattice structure}
	\label{sec:lattice}

The exact laws computed in Section~\ref{sec:algorithm} live on lattices whose holes, step
and residue classes are visible in every table; this section describes them.

\subsection{The uniform null}

The support and the parity of $d$ are given in~\cite[Proposition~2.1]{banic_elezovic_p07}:
for $n\ge4$ the only hole in $\{0,\ldots,2N\}$ is $d=1$ if $n\mid N$ and $d=0$ if
$n\nmid N$; for $n=3$ the values above $N$ form the band $k\equiv2N\pmod3$; for $n=2$,
$d\equiv N\pmod 2$; and $\Pr(d\text{ odd})=\frac12\bigl(1-(1-4/n)^N\bigr)$, so the
parity is exactly balanced only for $n=4$. The masses next to the holes are tiny in the
cases below (but not for very small $N$: at $(N,n)=(1,4)$ the hole is $d=0$ and
$\Pr(d=1)=\frac12$). Since
$d=0$ forces $X_1=\cdots=X_n$,
\[
	\Pr(d=0)=\frac{N!}{((N/n)!)^n\,n^N}\qquad(n\mid N),
\]
which is $4.9\cdot10^{-7}$, $2.4\cdot10^{-8}$, $1.1\cdot10^{-9}$ and $1.1\cdot10^{-27}$ at
$(N,n)=(50,10)$, $(100,10)$, $(200,10)$ and $(100,50)$; at $(101,10)$ the hole is at $0$
and $\Pr(d=1)=4.3\cdot10^{-8}$. For $n\ge4$ the holes therefore sit in the far left tail and do not affect the
upper-tail tests. The parity deviation $\frac12(1-4/n)^N$ is
$4.0\cdot10^{-12}$ at $(50,10)$ but $1.2\cdot10^{-4}$ at $(100,50)$.

\subsection{The lattice of the centered statistic}

\begin{theorem}[Lattice of $d_{\mathbf r}$]\label{thm:lattice}
Let $n\ge3$, $r_i=q_i/L$ with integers $q_i\ge1$, $\gcd_iq_i=1$, $\delta_i=q_{i+1}-q_i$ and
$V=L\,d_{\mathbf r}$. Put
\[
	g=\gcd\bigl(L,\;N\gcd_i\delta_i\bigr),\qquad
	s=\begin{cases}g,& L/g\ \text{odd},\\ 2g,& L/g\ \text{even},\end{cases}
\]
with $\gcd_i\delta_i=0$ if all $\delta_i=0$.
\begin{enumerate}
\item[(a)] $V$ is a non-negative integer, and $d_{\mathbf r}=0$ if and only if $X=N\mathbf r$.
Hence $\Pr(d_{\mathbf r}=0)>0$ if and only if $L\mid N$, and then it is the multinomial
probability of the point $N\mathbf r$.
\item[(b)] (Parity.) $V\equiv L(X_n-X_1)-N(q_n-q_1)\pmod 2$. If $L$ is even, the parity of
$V$ is that of $N(q_n-q_1)$. If $L$ is odd,
\[
	\Pr(V\ \text{even})=\tfrac12\Bigl(1+(-1)^{N(q_n-q_1)}\bigl(1-2(r_1+r_n)\bigr)^N\Bigr).
\]
\item[(c)] (Step.) The support of $V$ lies in one coset of $s\mathbb Z$. If
$N\ge N_0:=(4n-2)L/\min_iq_i$, the differences of support points generate exactly
$s\mathbb Z$, so $d_{\mathbf r}$ lives on a lattice of step exactly $s/L$.
\item[(d)] (Maximum.) With $\Delta^+=\max(\Delta,0)$, $\Delta^-=\max(-\Delta,0)$,
\[
	\max d_{\mathbf r}=N\Bigl(\sum_{i=1}^{n-1}|\Delta_i|+\max_{1\le j\le n}\mu_j\Bigr),\qquad
	\mu_j=\mathbf 1\{j\ge2\}(1-2\Delta_{j-1}^+)+\mathbf 1\{j\le n-1\}(1-2\Delta_j^-),
\]
attained at a vertex $X=N\mathbf e_j$. Under the uniform null this is $2N$.
\end{enumerate}
\end{theorem}

\begin{proof}
Write $y_i=L D_i-N\delta_i$, so that $V=\sum_i|y_i|$.

(a) $d_{\mathbf r}=0$ means $D_i=N\Delta_i$ for all $i$, that is
$X_i=X_1+N(r_i-r_1)$; summing over $i$ gives $X_1=Nr_1$, hence $X=N\mathbf r$. This
point has integer coordinates if and only if $L\mid Nq_i$ for all $i$, that is
$L\mid N\gcd_iq_i=N$.

(b) Since $|y|\equiv y\pmod 2$, $V\equiv\sum_iy_i=L(X_n-X_1)-N(q_n-q_1)$. For $L$ odd,
$V\equiv X_1+X_n+N(q_n-q_1)$, and $X_1+X_n\sim\Bin(N,r_1+r_n)$ is even with probability
$\frac12(1+(1-2(r_1+r_n))^N)$.

(c) \emph{Inclusion.} Each $y_i$ is a multiple of $\gcd(L,N\delta_i)$, hence of $g$, so
$g\mid V$. Moreover $N(q_n-q_1)=N\sum_i\delta_i$ is a multiple of $g$, and as in~(b)
\[
	V/g\equiv(L/g)(X_n-X_1)-N(q_n-q_1)/g\pmod 2 .
\]
If $L/g$ is even, $V/g$ has a fixed parity, so $V$ lies in a coset of $2g\mathbb Z$;
otherwise in $g\mathbb Z$.

\emph{Equality.} Let $N\ge N_0$. For a sign pattern $\varepsilon\in\{\pm1\}^{n-1}$ let
$w_i=\sum_{k<i}\varepsilon_k$, let $\bar w$ be their mean, and let $x$ be an integer vector
with $\sum x_i=N$ and $|x_i-x'_i|<1$, where $x'=N\mathbf r+4(w-\bar w\mathbf 1)$ (round down
and add $1$ to the coordinates with the largest fractional parts). The $w_i$ lie in an interval
of length at most $n-1$ that contains $\bar w$, so $|w_i-\bar w|\le n-1$ and
$x_i>Nr_i-4(n-1)-1\ge1$; and $D_i(x)=N\Delta_i+4\varepsilon_i+\eta_i$ with $|\eta_i|<2$,
so $\varepsilon_iy_i(x)>2L$. So every sign pattern is realised by a point of the support
whose edges are at distance more than $2L$ from $0$. Two consequences:
\begin{itemize}
\item Take $\varepsilon_1=\varepsilon_2$ and move one observation from cell $1$ to cell $2$.
Then $y_1$ changes by $2L$, $y_2$ by $-L$, no sign changes, and $V$ changes by
$\varepsilon_1(2L)-\varepsilon_2L=\varepsilon_1L$. So $L$ is a difference of support points.
\item On a point with sign pattern $\varepsilon$, $V=\sum_i\varepsilon_iy_i\equiv
-N\sum_i\varepsilon_i\delta_i\pmod L$. For patterns $\varepsilon,\varepsilon'$ that differ
only at the edge $i$, the difference of the two values of $V$ is $\equiv\pm2N\delta_i\pmod L$.
\end{itemize}
Hence the group generated by the differences of support points contains $L$ and
$2N\delta_i$ for every $i$, and so contains $\gcd(L,2N\gcd_i\delta_i)$. Write
$M=N\gcd_i\delta_i$; since $\gcd(L/g,M/g)=1$,
$\gcd(L,2M)=g\gcd(L/g,2M/g)=g\gcd(L/g,2)=s$. With the inclusion this proves~(c).

(d) $d_{\mathbf r}$ is a convex function of $x$ on the simplex
$\{x\ge0,\sum x_i=N\}$, so its maximum over the simplex is attained at a vertex
$N\mathbf e_j$, which is a support point; hence it is also the maximum over the
integer points. At $N\mathbf e_j$ only $D_{j-1}=N$ (if $j\ge2$) and $D_j=-N$ (if
$j\le n-1$) are non-zero, so, since $|\Delta_i|<1$,
$d_{\mathbf r}(N\mathbf e_j)=N\bigl(\sum_i|\Delta_i|+\mathbf 1\{j\ge2\}(1-\Delta_{j-1}-|\Delta_{j-1}|)
+\mathbf 1\{j\le n-1\}(1+\Delta_j-|\Delta_j|)\bigr)$, which is the formula.
\end{proof}

When the support of $V$ has at least two points we write $H$ for its actual span, the gcd
of the differences of its support points; by~(c), $s\mid H$, and $H=s$ for $N\ge N_0$. (A
one-point support, such as that of $\mathbf q=(1,2,1)$ at $N=1$, where $V=6$ for every
sample, has no step.)

So the step is proved to be exactly $s/L$ for $N\ge N_0$, for example for $N\ge40$ when
$\mathbf q=(1,2,1)$ and for $N\ge198$ when $\mathbf q=(1,3,2,4,1)$. Below $N_0$ only the
inclusion is proved, and the equality can fail when a cell has a small expected count: for
$\mathbf q=(1,8,1)$ and $N=3$ the support of $V$ is $\{12,18,42,72\}$, so $d_{\mathbf r}$ has
step $\frac35$ while $s/L=\frac15$; for $\mathbf q=(1,3N-1,1)$ the support of $V$ has span
$3\gcd(4,N-1)$ for every $N\ge3$ (with $J=X_1+X_3$, $V=6N$ if $J=0$ and
$V=3\bigl((3N+1)J-2N\bigr)$ if $J\ge1$, since then $X_2\le N-1$, so that
$X_2-X_1\le N-1$ and $X_3-X_2\ge-(N-1)$, and therefore the first centered edge
$L(X_2-X_1)-N(3N-2)\le-1$ is negative and the second $L(X_3-X_2)+N(3N-2)\ge1$ positive; hence $H=3\gcd(3N+1,N-1)=3\gcd(4,N-1)$), while
$s\in\{1,2\}$; and there are exceptions for $n=4$ and
$5$ as well, such as $\mathbf q=(1,13,1,1)$ and $\mathbf q=(1,1,12,1,1)$ at $N=3$. On our tables
the equality holds: by exhaustive enumeration for the $18$ profiles $(1,1,1)$, $(1,2,1)$,
$(1,3,1)$, $(1,2,3)$, $(2,1,2)$, $(1,4,1)$, $(4,1,4)$, $(1,1,2)$, $(1,1,1,1)$,
$(1,2,1,2)$, $(1,2,3,4)$, $(3,1,2,1)$, $(1,1,1,2)$, $(2,3,2,3)$, $(1,1,1,1,1)$,
$(1,2,1,2,1)$, $(1,2,3,4,5)$, $(1,3,2,4,1)$ and $1\le N\le12$ (respectively $N\le8$ for
$n=5$), except for $N\le2$: $\mathbf q=(1,2,1)$ at $N=2$ ($H=8$, $s=4$), $(1,3,1)$,
$(2,1,2)$ and $(4,1,4)$ at $N=1$, and $(1,2,1)$ at $N=1$, whose support is one point; and
directly from the
support on all $270$ exact tables of Section~\ref{sec:calibration} ($22$ profiles, $n\le12$,
$10\le N\le250$). Below $N_0$ we therefore use the actual span $H$ read from the exact support.
For $n=2$ the proof of~(c) fails (it moves an observation between two cells with equal edge
signs): there $V=2|Lx_1-Nq_1|$, whose step is $2\gcd(L,2N)=2g$ for $N\ge2$, that is $2s$ when $L/g$ is odd and $s$ when
$L/g$ is even (for $N=1$ the two values are $2q_1$ and $2q_2$); for the uniform null this is the
deterministic parity of~\cite[Proposition~2.1(c)]{banic_elezovic_p07}.

Table~\ref{tab:steps} shows how the step depends on $N$ modulo $L$. For
$\mathbf q=(1,2,1)$ the step is $1$ for even $N\ge4$, so
$d_{\mathbf r}$ then lives on the integers like the uniform $d$, and $\frac12$ for odd $N\ge3$ (at $N=2$ it is $2$, and
at $N=1$ the support is one point). Thus a continuity correction of
$1/(2L)$, half of the nominal step $1/L$, is wrong in general; Section~\ref{sec:calibration}
uses half of the true step.

\begin{table}[ht]
\centering\small
\caption{Lattice of $d_{\mathbf r}$ for six profiles (Theorem~\ref{thm:lattice}): the
actual span $H/L$ (``step''), read from the exact support, the smallest value, and the
mass at zero, from the exact laws (for the last three profiles at $N=101$ from the exact
support).}
\label{tab:steps}
\begin{tabular}{lrcccccc}
\toprule
 & & \multicolumn{3}{c}{$N=100$} & \multicolumn{3}{c}{$N=101$}\\
\cmidrule(lr){3-5}\cmidrule(lr){6-8}
$\mathbf q$ & $L$ & step & $\min d_{\mathbf r}$ & $\Pr(d_{\mathbf r}=0)$ & step & $\min d_{\mathbf r}$ & $\Pr(d_{\mathbf r}=0)$\\
\midrule
$(1,2,1)$ & 4 & $1$ & $0$ & $8.94\cdot10^{-3}$ & $\frac12$ & $\frac32$ & $0$\\
$(1,3,1)$ & 5 & $1$ & $0$ & $1.02\cdot10^{-2}$ & $\frac15$ & $\frac65$ & $0$\\
$(1,2,3)$ & 6 & $\frac13$ & $1$ & $0$ & $\frac13$ & $1$ & $0$\\
$(1,2,1,2,1)$ & 7 & $\frac17$ & $\frac{11}7$ & $0$ & $\frac17$ & $2$ & $0$\\
$(1,2,3,4,5)$ & 15 & $\frac13$ & $\frac43$ & $0$ & $\frac1{15}$ & $\frac{23}{15}$ & $0$\\
$(1,3,2,4,1)$ & 11 & $\frac1{11}$ & $\frac{19}{11}$ & $0$ & $\frac1{11}$ & $\frac{19}{11}$ & $0$\\
\bottomrule
\end{tabular}
\end{table}

\subsection{Three cells: a permanent sawtooth}

For $n=3$ under the uniform null the exact law is not unimodal: for every $N\ge3$ its
probability mass function on $\{0,\ldots,2N\}$ has at least two strict local maxima
(endpoints included; for $N\ge5$ the values $2N$ and $2N-3$ lie in the band
$k\equiv2N\pmod3$ above $N$ and so carry positive mass, while their neighbours $2N-1$,
$2N-2$ and $2N-4$ are holes; $N=3,4$ by direct count), and for $5\le N\le300$, by direct computation, at least two strict
local maxima in the interior of its support, each support point being compared with the
neighbouring support points; and it oscillates with period $3$~\cite[Numerical
observation~2.2]{banic_elezovic_p07}. The next theorem describes the oscillation. Let
$\varphi_v$ be the density of $\mathcal N(0,v)$ and put
\[
	f_{\mathrm{m}}(y)=2\varphi_{2/3}(y)\erf\Bigl(\frac y2\Bigr),\qquad
	f_{\mathrm{v}}(y)=3\varphi_2(y)\erf\Bigl(\frac{\sqrt3\,y}2\Bigr).
\]

\begin{theorem}[Local limit law for $n=3$]\label{thm:n3local}
Let $n=3$ and $\mathbf r=\mathbf u$. Uniformly in $k\ge0$, with $y=k/\sqrt N$,
\[
	\sqrt N\,\Pr(d=k)=f_{\mathrm m}(y)
	+f_{\mathrm v}(y)\bigl(\mathbf 1\{k\equiv N\}+\mathbf 1\{k\equiv-N\}\bigr)+O(N^{-1/2}),
\]
the congruences being modulo $3$. Thus if $3\mid N$ the values $k\equiv0$ carry the
density $f_{\mathrm m}+2f_{\mathrm v}$ and the others $f_{\mathrm m}$; if $3\nmid N$ the
classes $k\equiv\pm N$ carry $f_{\mathrm m}+f_{\mathrm v}$ and the class $k\equiv0$ carries
$f_{\mathrm m}$. In the limit, summed over
$k$, the $f_{\mathrm m}$ part has mass $\frac13$ and the $f_{\mathrm v}$ part, with its indicators, $\frac23$.
\end{theorem}

\begin{proof}
Let $A=X_3-X_1$ and $B=3X_2-N$, so that $d=\max(|A|,|B|)$
by Remark~\ref{rem:n3}. Here $X_2\sim\Bin(N,\frac13)$, and given
$X_2=m$, $A=M-2X_1$ with $M=N-m$ and $X_1\sim\Bin(M,\frac12)$. We use the local limit
theorem for the binomial law with its rate: $\Pr(\Bin(M,\pi)=j)$ equals the normal
density with the same mean and variance at $j$, up to $O(M^{-1})$ uniformly in $j$, for
fixed $\pi\in\{\frac13,\frac12\}$, and the Berry--Esseen bound
$\sup_x|\Pr(\Bin(M,\pi)\le x)-\Phi((x-M\pi)/\sqrt{M\pi(1-\pi)})|=O(M^{-1/2})$ for its
distribution function. Both are the case of i.i.d.\ Bernoulli summands (a lattice law of
span $1$ with moments of all orders) of standard results: the local one is the lattice
asymptotic expansion in the local limit theorem~\cite[Ch.~VII]{petrov1975} with one
correction term: that term is itself $O(M^{-1})$ uniformly in $j$, and so is the remainder,
so the normal density alone is accurate to $O(M^{-1})$; the global one is
the Berry--Esseen inequality~\cite[Ch.~V]{petrov1975}. All error terms below are uniform in $k$; values of $m$ with $|m-N/3|>N/6$ have
exponentially small probability and are absorbed in the errors. Split
\[
	\Pr(d=k)=\Pr(|B|=k,|A|<k)+\Pr(|A|=k,|B|<k)+\Pr(|A|=|B|=k).
\]
\emph{Ties.} $|B|=k$ fixes at most two values of $m$, each of probability $O(N^{-1/2})$, and
given $m$, $\Pr(|A|=k)=O(N^{-1/2})$; so the last term is $O(N^{-1})$.

\emph{Valleys and peaks} ($|B|=k$). $B=k$ requires $m=(N+k)/3$, i.e. $k\equiv-N\pmod3$, and
$B=-k$ requires $k\equiv N$. In either case the local limit theorem gives
$\Pr(X_2=m)=3N^{-1/2}\varphi_2(y)+O(N^{-1})$, and by Berry--Esseen
$\Pr(|A|<k\mid X_2=m)=\erf\bigl(k/\sqrt{2M}\bigr)+O(N^{-1/2})$ with
$M=(2N\mp k)/3$. Since $k/\sqrt{2M}=\frac{\sqrt3}2y\,(1\mp k/(2N))^{-1/2}$ and
$\varphi_2(y)\,y^2$ is bounded, replacing $k/\sqrt{2M}$ by $\frac{\sqrt3}2y$ costs
$O(N^{-1/2})$ after multiplication by $\varphi_2(y)$. This gives the term
$N^{-1/2}f_{\mathrm v}(y)$ for each admissible sign.

\emph{Monotone paths} ($|A|=k$, $k\ge1$). Given $m$, $A=\pm k$ requires $M\equiv k\pmod2$, and
then $\Pr(A=\pm k\mid X_2=m)=2M^{-1/2}\varphi_1(k/\sqrt M)+O(N^{-1})$ for each sign. Replacing
$M$ by $\frac23N$ changes this by $O(N^{-3/2}|m-\frac N3|)$, whose expectation over $X_2$
is $O(N^{-1})$. Hence
\[
	\Pr(|A|=k,|B|<k)=4N^{-1/2}\varphi_{2/3}(y)\,
	\Pr\bigl(|3X_2-N|<k,\ X_2\equiv N-k\ (\mathrm{mod}\ 2)\bigr)+O(N^{-1}).
\]
The two parity classes of $X_2$ differ in probability, on any interval, by at most the
total variation distance between $X_2$ and $X_2+1$, which is
$\max_k\Pr(X_2=k)=O(N^{-1/2})$ since the binomial law is unimodal; and
$\Pr(|3X_2-N|<k)=\erf(y/2)+O(N^{-1/2})$ by Berry--Esseen, since
$\Var(3X_2)=2N$. So this term is $N^{-1/2}f_{\mathrm m}(y)+O(N^{-1})$. For $k=0$,
$f_{\mathrm m}(0)=f_{\mathrm v}(0)=0$, and $d=0$ forces $X_1=X_2=X_3$, so that
$\Pr(d=0)=0$ if $3\nmid N$ and
$\Pr(d=0)=N!/\bigl((N/3)!^3\,3^N\bigr)=\frac{3\sqrt3}{2\pi N}\bigl(1+O(N^{-1})\bigr)$ by
Stirling's formula otherwise; hence $\sqrt N\Pr(d=0)=O(N^{-1/2})$.

Finally $f_{\mathrm m}$ integrates to $\Pr(|U|>|W|)$ for independent $U\sim\mathcal N(0,\frac23)$,
$W\sim\mathcal N(0,2)$, which is $\frac2\pi\arctan\frac1{\sqrt3}=\frac13$. Similarly
$\int_0^\infty f_{\mathrm v}=\frac32\Pr(|U|<|W|)=1$, and the indicator factor in the theorem
averages $\frac23$ over the three residue classes of $k$, so the $f_{\mathrm v}$ part has
mass $\frac23$.
\end{proof}

The sawtooth is permanent: the ratio of neighbouring masses near the mode tends to
$(f_{\mathrm m}+2f_{\mathrm v})/f_{\mathrm m}$, about $5.2$ when $3\mid N$. Table~\ref{tab:n3}
compares the theorem with the exact law (computed by the $O(N^2)$ formula of Remark~\ref{rem:n3}). The largest error over all $k$ of $\sqrt N\Pr(d=k)$ is $0.048$, $0.024$ and $0.017$ at
$N=300$, $1200$ and $2400$, attained at the atom $k=0$, where $\sqrt N\Pr(d=0)\approx3\sqrt3/(2\pi\sqrt N)$; over
$k\ge1$ it is $0.032$, $0.016$ and $0.011$, in line with the rate $N^{-1/2}$; at $N=301$ and $302$ it is $0.016$. The number of interior local maxima on the
support (endpoints excluded) above $10^{-3}$ of the mode is $30$, $61$ and $87$ at
$N=300$, $1200$, $2400$, consistent with growth like $\sqrt N$.

\begin{table}[ht]
\centering\small
\caption{$n=3$, uniform null: $\sqrt N\Pr(d=k)$, exact and by Theorem~\ref{thm:n3local}.}
\label{tab:n3}
\begin{tabular}{lccccccc}
\toprule
$N=300$, $k$ & 15 & 16 & 17 & 18 & 19 & 20 & 21\\
exact & 1.2509 & 0.2375 & 0.2558 & 1.2614 & 0.2137 & 0.2196 & 1.2073\\
limit & 1.2538 & 0.2506 & 0.2431 & 1.2634 & 0.2227 & 0.2106 & 1.2083\\
\midrule
$N=301$, $k$ & 13 & 14 & 15 & 16 & 17 & 18 & 19\\
exact & 0.7389 & 0.7350 & 0.2567 & 0.7651 & 0.7470 & 0.2348 & 0.7437\\
limit & 0.7306 & 0.7457 & 0.2561 & 0.7580 & 0.7558 & 0.2340 & 0.7374\\
\midrule
$N=2400$, $k$ & 45 & 46 & 47 & 48 & 49 & 50 & 51\\
exact & 1.2644 & 0.2442 & 0.2508 & 1.2681 & 0.2362 & 0.2409 & 1.2628\\
limit & 1.2647 & 0.2488 & 0.2462 & 1.2683 & 0.2402 & 0.2369 & 1.2631\\
\bottomrule
\end{tabular}
\end{table}

\subsection{Residue classes and orthant probabilities}

For a general null the lattice of Theorem~\ref{thm:lattice} carries a similar, coarser
structure. On a point where all $y_i=LD_i-N\delta_i$ are non-zero with signs
$\varepsilon\in\{\pm1\}^{n-1}$,
\[
	V=\sum_i\varepsilon_iy_i\equiv\rho_N(\varepsilon):=-N\sum_i\varepsilon_i\delta_i\pmod L,
\]
so the residue of $V$ modulo $L$ is a function of the sign pattern, as observed for the
cost of the recurrence in~\cite{banic_elezovic_gct}. Let $\mathbf G=\mathbf G^{(\mathbf r)}$
be the limit of $N^{-1/2}(\tilde D_1,\ldots,\tilde D_{n-1})$~\cite[Theorem~3.1(a)]{banic_elezovic_p07},
with the covariance of~\cite[Lemma~2.4]{banic_elezovic_p4}.

\begin{theorem}[Residue classes]\label{thm:residues}
Let $n\ge2$, $r_i=q_i/L$ with integers $q_i\ge1$ and $L=\sum_iq_i$, and $V=L\,d_{\mathbf r}$.
As $N\to\infty$,
\[
	\max_{\rho\in\mathbb Z/L\mathbb Z}\Bigl|\Pr(V\equiv\rho\ \mathrm{mod}\ L)
	-\sum_{\varepsilon:\,\rho_N(\varepsilon)=\rho}\Pr(\sgn\mathbf G=\varepsilon)\Bigr|\to0 .
\]
For $n=3$, $\Pr(\sgn G_1=\sgn G_2)=\frac12+\frac1\pi\arcsin\Corr(G_1,G_2)$.
\end{theorem}

\begin{proof}
The covariance of $\mathbf G$ is $\mathcal D\Sigma\mathcal D^{\top}$, with $\Sigma=\operatorname{diag}(\mathbf r)
-\mathbf r\mathbf r^{\top}$ and $\mathcal D$ the $(n-1)\times n$ difference matrix, $(\mathcal DX)_i=X_{i+1}-X_i$.
It is nonsingular (as also used in~\cite[proof of Theorem~4.10]{banic_elezovic_p07}): if
$y^{\top}\mathcal D\Sigma\mathcal D^{\top}y=0$, then $\mathcal D^{\top}y$ lies in the kernel of $\Sigma$, which is
spanned by $(1,\ldots,1)$ since all $r_i>0$; the coordinates of $\mathcal D^{\top}y$ sum to zero, so
$\mathcal D^{\top}y=0$, and $y=0$ since $\mathcal D$ has rank $n-1$. So the boundaries $\{G_i=0\}$ of the orthants are null sets, and the central limit
theorem gives $\Pr(\sgn\tilde D=\varepsilon)\to\Pr(\sgn\mathbf G=\varepsilon)$ for each of the
finitely many $\varepsilon$, while $\Pr(\tilde D_i=0)\to0$. On the event that $\tilde D$ has
pattern $\varepsilon$, $V\equiv\rho_N(\varepsilon)$. Summing over $\varepsilon$ gives the claim
uniformly in $\rho$, although the map $\rho_N$ changes with $N$. The case $n=3$ is
Sheppard's formula~\cite{sheppard1899}.
\end{proof}

The limit is periodic in $N$ with period $L$, through $\rho_N$. For
$\mathbf q=(1,2,1)$ and odd $N$, the class $V\equiv0\pmod4$ is the event of equal signs,
with limit $\frac12+\frac1\pi\arcsin(-\frac7{11})=0.2804$; the exact class mass is
$0.2778$, $0.2798$, $0.2803$ at $N=101$, $401$, $1601$. For $\mathbf q=(1,2,3)$ and
$N\equiv4\pmod6$ the three occupied classes have limits $0.6991$, $0.1505$, $0.1505$ and
exact masses $(0.7008,0.1392,0.1600)$, $(0.6995,0.1450,0.1554)$, $(0.6992,0.1478,0.1530)$ at
$N=100$, $400$, $1600$; the largest error halves when $N$ is multiplied by $4$. This is the
general rate: $\tilde D$ is a sum of $N$ independent bounded lattice vectors with nonsingular
covariance, and orthants are convex, so the multivariate Berry--Esseen bound for convex
sets~\cite{bhattacharya_rao1976,bentkus2003}, together with $\Pr(\tilde D_i=0)=O(N^{-1/2})$, gives an
error $O(N^{-1/2})$ in Theorem~\ref{thm:residues}, uniformly in $\rho$. For
$n=5$ the orthant probabilities are four-dimensional (we computed them by Gauss--Legendre
quadrature in double precision with $40$ and $56$ nodes per coordinate on each orthant,
truncated at $10$ standard deviations, with the last coordinate integrated in closed form;
the two orders agree to $10^{-14}$, and the randomized quasi-Monte Carlo routine of
Genz~\cite{genz1992} (SciPy~1.15.3, absolute and relative tolerance $10^{-9}$, at most
$10^7$ points; no user seed is involved, since its internal generator starts from a fixed
state, so that a fresh process repeats the run exactly), agrees to $10^{-7}$);
for $\mathbf q=(1,3,2,4,1)$ the
largest error over the nine occupied classes is $0.014$ at $N=100$ and $0.0034$ at
$N=250$. These errors are not worst cases: for $\mathbf q=(1,2,3,4,5)$ at $N=151$ the largest class
error is $0.043$ (exact $0.2518$ against the limit $0.2084$), so at these $N$ the orthant limit is a rough
guide to the class masses. The class masses are very unequal (from $0.007$ to $0.23$ in the last example),
so the exact law of $d_{\mathbf r}$ is a superposition of sub-lattice laws with different
weights. This is a mechanism analogous to the sawtooth of Theorem~\ref{thm:n3local}, but
coarser: under the uniform
null with $n=3$, $V=3d\equiv0\pmod 3$, and Theorem~\ref{thm:residues} does
not see the period-$3$ oscillation of $d$.

\section{Exact cumulants as a check of the analytic limits}
	\label{sec:cumulants}

The exact laws give exact cumulants at every tabulated $N$; here they are compared with
the analytic limits. By~\cite[Theorem~3.1(b)]{banic_elezovic_p07},
$d_{\mathbf r}/\sqrt N\to Q_{\mathbf r}=\sum_i|G_i|$
with convergence of all moments, and the third cumulant of $Q_{\mathbf r}$ is given in
closed form for every null in~\cite[Theorem~4.10]{banic_elezovic_p07}; for $n=3$ the limit
distribution function is a bivariate normal rectangle
probability~\cite[Proposition~4.14]{banic_elezovic_p07}, a product of error functions when $r_1=r_3$, and for
$n\le4$ all four limiting cumulants are elementary~\cite[Section~4]{banic_elezovic_p07}.
Under the uniform null
the analytic limits were compared with extrapolated exact values
in~\cite[Table~5]{banic_elezovic_p07}. Table~\ref{tab:cumulants} does the same for six
non-uniform nulls, using the exact tables of $d_{\mathbf r}$ for $N\le250$.

\begin{table}[ht]
\centering\small
\caption{Skewness $\gamma_1$ and excess kurtosis $\gamma_2$ of $d_{\mathbf r}$: the
analytic limit~\cite[Theorem~4.10, Proposition~4.14]{banic_elezovic_p07} (for $\gamma_2$
with $n=5$ a numerical evaluation of the limit law, see Numerical
observation~\ref{obs:cumulants}), the exact value at
$N=250$ (one of the fitted values in the first two rows, a value outside the fitted class in the last four), and the
extrapolation in $1/N$ of the exact values at the listed $N$, which all lie in one residue
class of $N$ modulo $L$ (least squares $a+b/N+c/N^2$, which is an exact
interpolation, with no residual degrees of freedom, when three values are used; $a+b/N$
through two values of $N$). The classes used are $N\equiv2\pmod4$, $0\pmod5$,
$2\pmod6$, $4\pmod7$, $5\pmod{15}$ and $2\pmod{11}$, in the order of the rows. Extrapolations from two or three
values of $N$ have no internal check and are printed to four decimals, about their accuracy
(Numerical observation~\ref{obs:cumulants}).}
\label{tab:cumulants}
\setlength{\tabcolsep}{3pt}\footnotesize
\begin{tabular}{lccccccc}
\toprule
$\mathbf q$ & $N$ used & $\gamma_1$ limit & $\gamma_1$, $N=250$ & $\gamma_1$ extrap.
 & $\gamma_2$ limit & $\gamma_2$, $N=250$ & $\gamma_2$ extrap.\\
\midrule
$(1,2,1)$ & $50,70,90,150,250$ & 1.101564 & 1.098676 & 1.101561 & 1.407483 & 1.389876 & 1.407484\\
$(1,3,1)$ & $50,60,\ldots,100,150,200,250$ & 1.120388 & 1.113475 & 1.120382 & 1.401374 & 1.379204 & 1.401383\\
$(1,2,3)$ & $50,80,200$ & 1.074517 & 1.072817 & 1.0745 & 1.371146 & 1.363821 & 1.3711\\
$(1,2,1,2,1)$ & $60,200$ & 0.760882 & 0.757685 & 0.7609 & 0.644111 & 0.632864 & 0.6443\\
$(1,2,3,4,5)$ & $50,80,200$ & 0.769071 & 0.766518 & 0.7691 & 0.731816 & 0.721669 & 0.7318\\
$(1,3,2,4,1)$ & $90,200$ & 0.762612 & 0.759929 & 0.7626 & 0.602361 & 0.594254 & 0.6024\\
\bottomrule
\end{tabular}
\end{table}

\begin{observation}[Approach to the limit]\label{obs:cumulants}
For the six nulls of Table~\ref{tab:cumulants} the exact skewness lies below its limit at
every tabulated $N$, and for the three nulls with $n=3$ at every $N$ with $50\le N\le255$
(computed from the exact law of Remark~\ref{rem:n3}); there $N(\gamma_1^\infty-\gamma_1)$ stays
between $0.30$ and $1.92$. The data are consistent with an approach of order $1/N$, as under the uniform
null~\cite[Numerical observation~5.1]{banic_elezovic_p07}, but, unlike there (where for $n=3$ the residue of $N$ enters
only at order $N^{-2}$), with a coefficient that depends on $N$ modulo $L$: for $\mathbf q=(1,2,1)$ it is about $0.72$--$0.75$ for
$N\equiv2\pmod4$, $0.96$--$0.99$ for odd $N$ and $1.84$--$1.92$ for $4\mid N$; for
$(1,2,3)$ it is about $0.30$--$0.33$, $0.41$--$0.43$, $0.80$--$0.82$ and $1.52$--$1.60$ for
$N\equiv3$, $N\equiv\pm2$, $N\equiv\pm1$ and $N\equiv0\pmod6$. Consequently an extrapolation in
$1/N$ is meaningful only within one residue class of $N$ modulo $L$. For $n=3$, a least-squares
fit $a+b/N+c/N^2$ to all $N$ in $[50,250]$ of the class used in
Table~\ref{tab:cumulants} ($51$, $41$ and $34$ values) reproduces the
analytic limits of $\gamma_1$ and $\gamma_2$ to within $10^{-5}$, while $a+b/N$ is off by up
to $7\cdot10^{-4}$, so the $1/N^2$ term matters. Table~\ref{tab:cumulants} uses fewer values:
with three values of $N$ the quadratic fit is an exact interpolation (a Richardson
extrapolation with no internal check), and it reproduces the limits of $\gamma_1$ (and, for
$n=3$, of $\gamma_2$) to within $10^{-5}$ in all three classes of $(1,2,3)$ and of
$(1,2,3,4,5)$ represented on our grid; with two values of $N$ the linear fit is off by less
than $6\cdot10^{-5}$ in $\gamma_1$. For the five profiles whose
grid meets several classes, a fit over the whole grid, which mixes the classes, is off
by $5\cdot10^{-4}$ to $1.4\cdot10^{-2}$ (for example by $8.6\cdot10^{-3}$ in $\gamma_1$ for
$(1,2,1)$).
For $n\ge5$ we do not know a general elementary form of the limiting $\gamma_2$ under a
non-uniform null, but it
is computable from four-variate absolute normal moments (Nabeya~\cite{nabeya1961},
Kan and Robotti~\cite{kanrobotti2017}; see~\cite[Section~4]{banic_elezovic_p07}). We
evaluated it by Gauss--Legendre quadrature of the Gaussian limit, conditioning on one edge
after the other and splitting at the kinks; the values in Table~\ref{tab:cumulants} are stable
to $10^{-12}$ under a change of the number of nodes ($32$ and $48$ nodes per conditioning
step), and the same computation reproduces
every other limit of the table. The extrapolations differ from them by $1.9\cdot10^{-4}$
(two values of $N$), $3\cdot10^{-6}$ and $1.8\cdot10^{-5}$; for $(1,2,3,4,5)$ the three residue
classes give $0.731816$, $0.731813$ and $0.731815$.
\end{observation}

The means and variances agree with the limits in the same way: for example, for
$(1,3,1)$ at $N=250$, $\E d_{\mathbf r}/\sqrt N=1.276299$ and $\Var d_{\mathbf r}/N=0.666622$
against $\E Q_{\mathbf r}=1.276615$ and $\Var Q_{\mathbf r}=0.666675$, the latter being the
variance coefficient of~\cite[Corollary~6.4]{banic_elezovic_p4}. The large-$n$ profile law
of~\cite[Theorem~4.11]{banic_elezovic_p07} concerns smooth profiles with $n\to\infty$ and
cannot be tested on the tables of this paper (fixed $n\le12$, non-smooth weights).

All limits in this section, and the calibration of Section~\ref{sec:calibration}, concern
a fixed $n$ and a fixed null with all $r_i>0$, as $N\to\infty$. They say nothing about
sparse regimes in which $n$ grows with $N$ or some expected counts $Nr_i$ stay bounded.
Under the uniform null with $N^2/n\to0$, for example, with probability tending to one no
cell holds two observations, no two occupied cells are adjacent and the end cells are
empty; then every occupied cell contributes $2$ and $d=2N$, far from the $\sqrt N$ scale.
The example at the end of Section~\ref{sec:calibration} shows that moment-based
calibration can fail completely when expected counts are small.

\section{Calibration of continuous approximations}
	\label{sec:calibration}

With the exact laws of Section~\ref{sec:algorithm} and the lattice of
Section~\ref{sec:lattice} at hand, continuous approximations can be judged on the lattice,
case by case. In this section a \emph{cell} of a size table is a pair (case, level), not a
cell of the histogram.

\subsection{Measures}

In this section $Y$ is the statistic being calibrated: $Y=d$ under the uniform null and
$Y=d_{\mathbf r}=V/L$ under a general null. The continuous approximations below need a
support with at least two points and $\sigma>0$; a degenerate law (such as
$\mathbf q=(1,2,1)$ at $N=1$, or uniform $n=2$ at $N=1$) goes directly to the exact
procedure. Let $F_N(x)=\Pr(Y\le x)$ be its exact
distribution function, $\pi_k=\Pr(Y=k)$ the exact masses on the support, $h$ the lattice step ($h=1$ for $d$ with $n\ge3$, and $h=2$ for $d$ with $n=2$ and $N\ge2$, since then $d=|N-2X_1|$; $h=H/L$ for $d_{\mathbf r}$, with
$H$ the gcd of the support differences; $H=s$ by Theorem~\ref{thm:lattice}(c) for
$N\ge N_0$, and we checked $H=s$ directly on all $270$ tables below, although only $60$ of
the $182$ subgrid tables have $N\ge N_0$) and $G$ a continuous approximation. Following
\cite[Section~7.1]{banic_elezovic_p07} we compare $F_N(k)$ with $G(k+\frac h2)$, half a
lattice step above $k$ (the midpoint to the next lattice site), and use, over the support
points $k$,
\[
	\KS=\max_k\bigl|F_N(k)-G(k+\tfrac h2)\bigr|,\qquad
	\AD=\sum_{k:\,F_N(k)<1}\pi_k\frac{(F_N(k)-G(k+\frac h2))^2}{F_N(k)(1-F_N(k))},
\]
and $\LL=\bigl(\sum_k(F_N(k)-G(k+\frac h2))^2\bigr)^{1/2}$. ($\AD$ uses the exact law as
reference measure; it is a descriptive distance, not the Anderson--Darling test statistic.)
The uncorrected sup-norm
$\sup_x|F_N(x)-G(x)|$ is not used: it is at least half the largest atom, that is at least
$0.030$, $0.021$, $0.015$, $0.020$ and $0.014$ in the five uniform cases below, and so it
measures the lattice rather than the fit. A continuous approximation with $(1-\alpha)$-quantile
$x$ gives the critical value by the \emph{half-span rule}
of~\cite[Section~7.3]{banic_elezovic_p07}: $c$ is the smallest support point with $c\ge x+\frac h2$ ($c=+\infty$, never reject, if
there is none). The half-span rule is the simplest lattice correction. Expansions of
lattice laws with periodic (discontinuous) terms go back to Esseen~\cite{esseen1945} (see
Kolassa and McCullagh~\cite{kolassa_mccullagh1990} for the continuity-corrected Edgeworth
series); for sums of lattice random vectors Yarnold~\cite{yarnold1972} gives such an expansion of
the probability of a convex set (applied there to $X^2$; it also covers
$\{d_{\mathbf r}\le c\}$, a convex set of the edge vector), and for multinomial statistics such terms are also given by Siotani and Fujikoshi~\cite{siotani1984}; for the DTV they would have to be summed over the residue
classes of Theorem~\ref{thm:residues}. The half-span rule is therefore a rounding
convention, not a theorem: a single global shift does not represent the unequal residue
weights, and only an exact or
certified tail guarantees the level.

The exact critical value $c^*$ is the smallest support point with
$\Pr(Y\ge c^*)\le\alpha$, and the \emph{size} of a rule is $\Pr(Y\ge c)$. If no support
point qualifies, put $c^*=+\infty$: the test never rejects. This exact test is
non-randomized and has size at most $\alpha$, which can lie well below $\alpha$ when the
step is coarse. In terms of $p$-values, for the observed value $t$ of $Y$: $\Pr(Y\ge t)$
gives this test; the randomized value $\Pr(Y>t)+U\Pr(Y=t)$, with $U$ uniform on $(0,1)$ and
independent of $X$, is exactly uniform under the null. Equivalently, if $c^-$ is the
support point just below $c^*$ (the largest support point if $c^*=+\infty$), the
randomized test that rejects when $Y\ge c^*$, and with probability
$(\alpha-\Pr(Y\ge c^*))/\Pr(Y=c^-)$ when $Y=c^-$, has size exactly
$\alpha$~\cite{tocher1950}. The mid-$p$ value $\Pr(Y>t)+\frac12\Pr(Y=t)$~\cite{lancaster1961}
is less conservative but does not guarantee the level. Of the procedures compared here, only the exact
ones, the non-randomized and the randomized test, control the size; we report the size of the
non-randomized test throughout.

The approximations are the moment-matched gamma law ($\nu=\mu^2/\sigma^2$,
$\theta=\sigma^2/\mu$); the moment-matched beta law on $[0,c_B]$, in particular on
$[0,2N]$ and the law $B_3$ on $[0,c_3]$ with
$c_3=\bigl(4\sigma\mu+\gamma_1(\sigma^2-\mu^2)\bigr)/(2\sigma-\gamma_1\mu)$, the unique support
for which its skewness equals the exact skewness $\gamma_1$ (it exists if and only if
$\sigma/\mu-\mu/\sigma<\gamma_1<2\sigma/\mu$, as shown before Numerical
observation~\ref{obs:beta3}); the normal
Cornish--Fisher quantile $\mu+\sigma C(z_{1-\alpha};\gamma_1,\gamma_2)$; and the
gamma-based Cornish--Fisher quantile ($\Gamma$+CF)
\[
	x^{\Gamma+\mathrm{CF}}=x^\Gamma+\sigma\bigl[C(z_{1-\alpha};\gamma_1,\gamma_2)
	-C(z_{1-\alpha};2\nu^{-1/2},6\nu^{-1})\bigr]
\]
of~\cite[eq.~(16)]{banic_elezovic_p07}, where $z_{1-\alpha}=\Phi^{-1}(1-\alpha)$,
\[
	C(z;g_1,g_2)=z+\frac{g_1}{6}(z^2-1)+\frac{g_2}{24}(z^3-3z)-\frac{g_1^2}{36}(2z^3-5z)
\]
is the second-order Cornish--Fisher polynomial~\cite[eq.~(15)]{banic_elezovic_p07},
$x^\Gamma$ is the $(1-\alpha)$-quantile of the moment-matched gamma law, and
$\mu,\sigma,\gamma_1,\gamma_2$ are the exact moments.

\subsection{The uniform null}

We use five illustrative cases, $(N,n)=(50,10)$, $(100,10)$, $(200,10)$, $(100,50)$,
$(200,50)$, which show moderate and large $n$; they are not a designed sample, and the full
uniform grid is calibrated in~\cite[Section~7.4, Table~12]{banic_elezovic_p07}.
Table~\ref{tab:uniform-fit} gives the continuity-corrected measures and
Table~\ref{tab:uniform-size} the sizes.

\begin{table}[ht]
\centering\small
\caption{Uniform null: continuity-corrected distances of the gamma law, the moment-matched
beta law on $[0,2N]$, and the beta law $B_3$ whose support $c_3$ matches the exact
skewness (Numerical observation~\ref{obs:beta3}).}
\label{tab:uniform-fit}
\begin{tabular}{lccccccccc}
\toprule
 & \multicolumn{3}{c}{$\KS$} & \multicolumn{3}{c}{$\AD$} & \multicolumn{3}{c}{$\LL$}\\
\cmidrule(lr){2-4}\cmidrule(lr){5-7}\cmidrule(lr){8-10}
$(N,n)$ & $\Gamma$ & $B[0,2N]$ & $B_3$ & $\Gamma$ & $B[0,2N]$ & $B_3$ & $\Gamma$ & $B[0,2N]$ & $B_3$\\
\midrule
$(50,10)$ & 0.0090 & 0.0031 & 0.0018 & $2.3\cdot10^{-4}$ & $2.7\cdot10^{-5}$ & $7.7\cdot10^{-6}$ & 0.028 & 0.0099 & 0.0052\\
$(100,10)$ & 0.0074 & 0.0011 & 0.0011 & $1.6\cdot10^{-4}$ & $2.8\cdot10^{-6}$ & $2.8\cdot10^{-6}$ & 0.028 & 0.0038 & 0.0037\\
$(200,10)$ & 0.0066 & 0.0017 & 0.0007 & $1.3\cdot10^{-4}$ & $8.2\cdot10^{-6}$ & $1.2\cdot10^{-6}$ & 0.029 & 0.0074 & 0.0029\\
$(100,50)$ & 0.0068 & 0.0039 & 0.0006 & $1.2\cdot10^{-4}$ & $3.8\cdot10^{-5}$ & $1.1\cdot10^{-6}$ & 0.025 & 0.0143 & 0.0025\\
$(200,50)$ & 0.0048 & 0.0018 & 0.0004 & $5.7\cdot10^{-5}$ & $8.0\cdot10^{-6}$ & $4.0\cdot10^{-7}$ & 0.021 & 0.0079 & 0.0018\\
\bottomrule
\end{tabular}
\end{table}

\begin{table}[ht]
\centering\small
\caption{Uniform null: size divided by $\alpha$ with the half-span rule $c\ge x+\frac12$
(an asterisk marks $c=c^*$), and the size divided by $\alpha$ of the exact test. The last column uses
$\Gamma$+CF with $c=\lceil x\rceil$.}
\label{tab:uniform-size}
\begin{tabular}{llcccccc}
\toprule
$(N,n)$ & $\alpha$ & exact test & $\Gamma$ & $B[0,2N]$ & $B_3$ & $\Gamma$+CF & $\Gamma$+CF, $\lceil x\rceil$\\
\midrule
$(50,10)$ & $5\%$ & 0.96 & 0.75 & 0.96* & 0.96* & 0.96* & 0.96*\\
 & $1\%$ & 0.91 & 0.67 & 0.91* & 0.91* & 0.91* & 0.91*\\
 & $0.1\%$ & 0.89 & 0.43 & 1.27 & 0.89* & 0.89* & 0.89*\\
$(100,10)$ & $5\%$ & 0.89 & 0.89* & 0.89* & 0.89* & 0.89* & 1.06\\
 & $1\%$ & 0.96 & 0.78 & 0.96* & 0.96* & 0.96* & 0.96*\\
 & $0.1\%$ & 0.99 & 0.46 & 0.99* & 0.99* & 0.99* & 0.99*\\
$(200,10)$ & $5\%$ & 0.97 & 0.86 & 0.97* & 0.97* & 0.97* & 0.97*\\
 & $1\%$ & 0.98 & 0.84 & 0.84 & 0.98* & 0.98* & 0.98*\\
 & $0.1\%$ & 0.90 & 0.53 & 0.90* & 0.90* & 0.90* & 1.07\\
$(100,50)$ & $5\%$ & 0.98 & 0.81 & 0.98* & 0.98* & 0.98* & 0.98*\\
 & $1\%$ & 0.84 & 0.65 & 1.07 & 0.84* & 0.84* & 1.07\\
 & $0.1\%$ & 0.94 & 0.51 & 1.26 & 0.94* & 0.94* & 0.94*\\
$(200,50)$ & $5\%$ & 0.92 & 0.92* & 0.92* & 0.92* & 0.92* & 1.06\\
 & $1\%$ & 0.88 & 0.74 & 1.04 & 0.88* & 0.88* & 1.04\\
 & $0.1\%$ & 0.95 & 0.63 & 1.17 & 0.95* & 0.95* & 0.95*\\
\bottomrule
\end{tabular}
\end{table}

\begin{observation}[Uniform null]\label{obs:uniform}
On the five cases of Tables~\ref{tab:uniform-fit}--\ref{tab:uniform-size}:
\begin{enumerate}
\item[(a)] $\Gamma$+CF with the half-span rule and the exact moments gives the exact
critical value in all $15$ cells (with the limiting $\gamma_1,\gamma_2$
of~\cite[Theorem~4.2]{banic_elezovic_p07} in $10$ of them, the other five being
conservative); with $c=\lceil x\rceil$, $\Gamma$+CF
is exact in $10$ cells and anti-conservative in $5$ ($1.04$--$1.07\alpha$).
\item[(b)] With the continuity correction the beta law on $[0,2N]$ is closer to the exact law
than the gamma law on all three measures in all five cases, by factors from $1.7$ to $55$. In the
tail the gamma law is conservative ($0.43$--$0.63\alpha$ at $0.1\%$, $0.65$--$0.84\alpha$ at
$1\%$), while the beta law is anti-conservative in $5$ of the $15$ cells, up to $1.27\alpha$.
\end{enumerate}
\end{observation}

So on these five cases the gamma law is conservative (on the wider grid
of~\cite[Table~12]{banic_elezovic_p07} it is anti-conservative in $2\%$ of the cells at
$5\%$), and the beta law on $[0,2N]$ is
closer in the body but not safe in the tail. The next proposition explains why the beta
law is close at moderate $N$ and why the two laws cannot be distinguished as $N\to\infty$.

\begin{proposition}[Beta on {$[0,2N]$} versus gamma]\label{prop:beta-gamma}
Let $\mu,\sigma^2$ be a mean and a variance with $\mu(c_B-\mu)>\sigma^2$, and let $B$ be the
beta law on $[0,c_B]$ with these moments. With $m=\mu/c_B$ and $v=\sigma^2/c_B^2$,
\[
	\gamma_1(B)=\frac{2(1-2m)\sqrt v}{m(1-m)+v},\qquad
	R_B:=\frac{\gamma_1(B)}{2\sigma/\mu}=\frac{1-2m}{1-m+v/m}.
\]
For the uniform null at fixed $n$ and $c_B=2N$, $R_B\to1$ as $N\to\infty$, and the
moment-matched beta and gamma laws of $d/\sqrt N$ converge weakly to the same gamma law
with shape $(\E S)^2/\Var S$ and scale $\Var S/\E S$, where $S$ is the limit of $d/\sqrt N$.
\end{proposition}

\begin{proof}
The moment-matched beta has parameters $a=m\lambda$, $b=(1-m)\lambda$ with
$\lambda=m(1-m)/v-1$, and the skewness of a beta law is
$2(b-a)\sqrt{a+b+1}/((a+b+2)\sqrt{ab})$; substituting gives the formula, and dividing by
the gamma skewness $2\sqrt v/m$ gives $R_B$. For $c_B=2N$,
$\mu=\E d\sim\sqrt N\,\E S$ and $\sigma^2\sim N\Var S$~\cite[Theorem~3.1(b)]{banic_elezovic_p07},
so $m=O(N^{-1/2})$ and $v/m=\sigma^2/(c_B\mu)=O(N^{-1/2})$, hence $R_B\to1$. Moreover
$a=m^2(1-m)/v-m\to(\E S)^2/\Var S$ and $b\to\infty$, and $bY\to\Gamma(a,1)$ in law for
$Y\sim\mathrm{Beta}(a,b)$ with $a$ converging and $b\to\infty$. Since
$c_B/(b\sqrt N)\to\Var S/\E S$, the law of $c_BY/\sqrt N$ converges to the stated gamma
law, which is also the limit of the moment-matched gamma law of $d/\sqrt N$.
\end{proof}

At $n=10$ the factor $R_B$ is $0.69$, $0.80$, $0.86$ at $N=50$, $100$, $200$, against the
exact $R_3=\gamma_1/(2\sigma/\mu)$ of $0.75$, $0.80$, $0.82$; at $n=50$ it is $0.39$ and
$0.62$ against $0.60$ and $0.72$. So the beta law on $[0,2N]$ behaves like a gamma law whose skewness
is shrunk by an amount fixed by the mechanical choice $c_B=2N$ together with $\mu$ and
$\sigma$, not by the shape of the law. Near $(N,n)=(100,10)$ this
coincides with the true deficit $R_3<1$ of~\cite[Corollary~4.5]{banic_elezovic_p07}; at
$n=50$ the shrinkage overshoots, which explains the anti-conservative sizes. The support
should therefore be chosen by the skewness, not by the ceiling $2N$.

Let $B_3$ be the moment-matched beta law on $[0,c_3]$ with $c_3$ chosen so that
$\gamma_1(B_3)$ equals the exact skewness. By Proposition~\ref{prop:beta-gamma},
$\gamma_1(B)=2\sigma(c_B-2\mu)/(\mu c_B+\sigma^2-\mu^2)$, whose derivative in $c_B$ is
$2\sigma(\mu^2+\sigma^2)/(\mu c_B+\sigma^2-\mu^2)^2>0$;
$\gamma_1(B)$ tends to $\sigma/\mu-\mu/\sigma$ as $c_B$ decreases to the smallest admissible support $\mu+\sigma^2/\mu$
(which is excluded), and to the gamma skewness $2\sigma/\mu$ as $c_B\to\infty$. So $c_3$ exists if and only if
$\sigma/\mu-\mu/\sigma<\gamma_1<2\sigma/\mu$, and it is then unique,
\[
	c_3=\frac{4\sigma\mu+\gamma_1(\sigma^2-\mu^2)}{2\sigma-\gamma_1\mu},
\]
and it exists in particular whenever $0<\gamma_1<2\sigma/\mu$ and $\mu>\sigma$.

\begin{observation}[Skewness-fitted beta]\label{obs:beta3}
The two conditions $0<\gamma_1<2\sigma/\mu$ and $\mu>\sigma$ hold on all $182$ subgrid
tables below. They fail for the sparse examples at the end of this section,
but $B_3$ still exists for two of them ($c_3=11.66$ for $\mathbf q=(1,1,1,10000,1,1,1)$, $N=21$, and $c_3=10.18$ for
the uniform null at $(N,n)=(5,200)$); it does not exist for $\mathbf q=(1,1000,1)$, $N=30$, where
$\gamma_1=4.17>2\sigma/\mu=3.85$. In the five uniform cases
$c_3=119.6$, $202.1$, $318.0$, $267.4$, $503.9$, against $2N=100$, $200$, $400$, $200$,
$400$. With the half-span rule $B_3$ gives the exact critical value in all $15$ cells, and
it has the smallest $\KS$, $\AD$ and $\LL$ in $14$ of the $15$ pairs (case, measure); at
$(100,10)$ its $\KS$ is slightly larger than that of $B[0,2N]$ ($0.00107$ against $0.00106$)
(Tables~\ref{tab:uniform-fit}--\ref{tab:uniform-size}).
\end{observation}

The ceiling of the support is not a natural parameter of the beta law here: the largest value of $d$ ($2N$
for $n\ge3$) lies of order $\sqrt N$ standard deviations above the mean (about $1.9\sqrt N$ at $(N,n)=(200,10)$), and the
fixed-$n$ limit $S$ of $d/\sqrt N$ has a Gaussian extreme tail~\cite[Proposition~4.8]{banic_elezovic_p07}. For a non-uniform
null the exact ceiling is given by Theorem~\ref{thm:lattice}(d), but it plays no role in
the calibration either.

\subsection{A general null}

{\sloppy The exact tables of $d_{\mathbf r}$ (linear statistic, $V=L\,d_{\mathbf r}$) cover $22$
profiles:
$n=3$: $\mathbf q=(1,2,1)$, $(1,3,1)$, $(1,4,1)$, $(1,2,3)$;
$n=4$: $(1,2,1,2)$, $(1,2,3,2)$, $(1,2,3,4)$, $(1,4,2,3)$;
$n=5$: $(1,2,1,2,1)$, $(1,3,1,3,1)$, $(1,2,3,2,1)$, $(1,2,3,4,5)$, $(1,3,2,4,1)$, $(2,4,1,3,1)$;
$n=6$: $(1,2,1,2,1,2)$, $(3,1,2,4,1,2)$, $(3,1,4,1,5,2)$;
$n=7$: $(1,2,1,2,1,2,1)$; $n=10$: $(1,2)$ repeated;
$n=12$: $(1,2)$, $(1,2,3)$ and $(1,2,3,2)$ repeated;
with $N\in\{10,20,\ldots,100\}$ and also $N=150,200$ ($n\le7$) and $N=250$ ($n\le5$), $270$
tables in all. As in the regular subgrid of~\cite[Numerical observation~7.1]{banic_elezovic_p07}
(which in addition requires $n\ge5$) we use $N\ge50$, $N\ge3n$: $182$ tables, $546$ cells (table, level). The profiles were chosen to cover
several shapes for each $n$, not sampled from a population of nulls; all have moderate
expected counts ($\min_iNr_i\ge2.08$ on the subgrid).\par}

\begin{observation}[Calibration under a general null]\label{obs:general}
On the $182$ tables of the subgrid, with the exact moments of each table (an oracle
calibration: these moments come from the exact law itself, so it tests the shape of the
approximations, not a procedure that avoids the exact law):
\begin{enumerate}
\item[(a)] $\Gamma$+CF with the half-span rule (shift $H/(2L)$, equal to $s/(2L)$ on all
our tables) gives the exact critical value in $53\%$, $40\%$ and $39\%$ of the tables
($96$, $73$ and $71$ of $182$) at $\alpha=5\%$, $1\%$ and $0.1\%$, with
largest sizes $1.30\alpha$, $1.35\alpha$ and $1.49\alpha$; the size exceeds $1.1\alpha$ in
$7\%$, $7\%$ and $15\%$ of the tables ($12$, $12$ and $28$ of $182$).
\item[(b)] By $n$, over the three levels together: $n=3$: exact in $36$ of $108$ cells
($33\%$), largest size $1.49\alpha$; $n=4$: $65/108$ ($60\%$), $1.11\alpha$; $n=5$: $35/162$
($22\%$), $1.20\alpha$; $n=6$: $27/72$ ($38\%$), $1.09\alpha$; $n=7$: $20/24$ ($83\%$),
$1.04\alpha$; $n=10$: $18/18$ ($100\%$); $n=12$: $39/54$ ($72\%$), $1.03\alpha$.
\item[(c)] The shift matters but is not the cause of the failures. With no shift the rule is
anti-conservative in $77$--$86\%$ of the tables ($141$, $157$ and $150$ of $182$ at $5\%$,
$1\%$ and $0.1\%$), with the naive shift $1/(2L)$ in $68$--$80\%$ ($123$, $146$, $141$), and
with the shift $\frac12$ it is exact in only $37$--$45\%$ ($68$, $81$, $73$). The half-span
shift gives nearly the same results as the midpoint between consecutive occupied support
points (the chosen critical value differs in $7$, $5$ and $11$ of the $182$ tables; at
$0.1\%$ the largest size is $1.44\alpha$ instead of $1.49\alpha$). With it the rule is
anti-conservative in $45$--$57\%$ of the tables ($82$, $103$, $103$), that is, the $\Gamma$+CF quantile lies at least
half a step below the exact quantile; since the placement is (nearly) the natural
one, the
misses are not mainly due to the choice among global shifts.
\item[(d)] The skewness-fitted beta $B_3$ of Numerical observation~\ref{obs:beta3}, with the
same shift, is exact in $68\%$, $64\%$, $41\%$ of the tables ($123$, $117$, $74$ of $182$),
anti-conservative in $25\%$, $12\%$, $7\%$ ($45$, $22$, $13$), with largest sizes
$1.19\alpha$, $1.16\alpha$, $1.07\alpha$; at $n=3$ it is exact in $33$ of $108$ cells
($31\%$) with largest size $1.19\alpha$.
\item[(e)] When the step is coarse the exact test itself is crude: the exact size is below
$0.8\alpha$ in $59$ of the $546$ cells, and as low as $0.39\alpha$; a randomized test at
the boundary point would attain $\alpha$ exactly.
\item[(f)] The normal Cornish--Fisher quantile $\mu+\sigma C(z_{1-\alpha};\gamma_1,\gamma_2)$, with the
same exact moments and shift, is exact in $100$, $88$ and $102$ of the $182$ tables at $5\%$,
$1\%$ and $0.1\%$ (against $96$, $73$, $71$ for $\Gamma$+CF), anti-conservative in $75$, $87$ and
$54$ (against $82$, $103$, $103$), with largest sizes $1.30\alpha$, $1.35\alpha$ and $1.12\alpha$
(against $1.30\alpha$, $1.35\alpha$, $1.49\alpha$); the size exceeds $1.1\alpha$ in $12$, $12$ and
$5$ tables. So the two tie at $5\%$, and the normal version is better at $1\%$ and clearly
better at $0.1\%$. By $n$, over the three levels together, it is exact in $32/108$ cells for
$n=3$ (against $36$ for $\Gamma$+CF), $79/108$, $57/162$, $37/72$, $23/24$, $18/18$ and $44/54$ for
$n=4,5,6,7,10,12$. In the five uniform cases of Table~\ref{tab:uniform-size} it is exact in all
$15$ cells, like $\Gamma$+CF and $B_3$, so these cases do not separate the rules. The
moment-matched gamma law alone and the three-moment shifted gamma (Pearson type~III) are
conservative at $0.1\%$ (exact in $0$ and $6$ tables, largest sizes $0.71\alpha$ and $0.90\alpha$),
and the gamma law alone is exact in only $11$ tables at $1\%$.
\end{enumerate}
\end{observation}

Part~(c) is illustrated by $\mathbf q=(1,2,1)$, $N=200$ (step $1$): at $\alpha=1\%$ the
normal Cornish--Fisher test, and also $\Gamma$+CF, has size $1.31\alpha$ with the naive shift
$\frac18$ and $0.87\alpha$, the exact critical value, with the half-span shift $\frac12$;
but at $\alpha=5\%$ both shifts give $1.12\alpha$, against $0.80\alpha$ for the exact test.
At $N=211$ (step $\frac12$) the normal Cornish--Fisher test at $1\%$ has size $1.30\alpha$
with the naive shift and $0.87\alpha$ with the true one, while $\Gamma$+CF has $1.30\alpha$
even with the true shift. These values were obtained from the
exact law of Remark~\ref{rem:n3}; those for $N=200$ are also quoted
in~\cite[Section~7.3]{banic_elezovic_p07}.

\paragraph{Practical rule.}
The calibration above is a diagnostic, not a level guarantee: it uses the exact moments of
each table, which we read from the exact law, and its cases were chosen for illustration.
No approximation-only procedure is proposed here as a level-$\alpha$ test: whenever the
level must be controlled, an approximate critical value is followed by the exact tail check
described below.

\emph{Uniform null.} Here $\Gamma$+CF with the half-span rule was exact in all $15$ cells of
Table~\ref{tab:uniform-size} ($n=10,50$), and in $10$ of them with the limiting
$\gamma_1,\gamma_2$ of~\cite[Theorem~4.2]{banic_elezovic_p07}; on the wider grid
of~\cite[Section~7.4, Table~12]{banic_elezovic_p07} it is anti-conservative at small $n$, up
to $1.47\alpha$ at $n=2$ and $1.19\alpha$ at $n=3$ (at $(N,n)=(90,3)$ the size is
$1.14\alpha$ at $5\%$ and $1.11\alpha$ at $0.1\%$, with the exact and with the limiting
$\gamma_1,\gamma_2$ alike). Under the uniform null we therefore
follow the three tiers of~\cite[Section~7.4]{banic_elezovic_p07}, in which the corrected
quantile uses the limiting $\gamma_1,\gamma_2$ and the exact mean and variance: the exact
law for $n\le4$; for $n\ge5$ the half-span rule when an approximate critical value is
acceptable, and the
full-span rule $c\ge x+h$ as a more conservative option. Both
are observations on the grid $N\ge40$, $N\ge3n$ ($N\ge100$ at $\alpha=0.1\%$)
of~\cite[Section~7.4]{banic_elezovic_p07}, not level guarantees, and outside it the
$\Gamma$+CF quantile with the exact moments can fail badly: for $(N,n)=(5,200)$ and
$\alpha=0.1\%$ the exact $\gamma_1=-2.65$ and $\gamma_2=6.56$ give
$x^{\Gamma+\mathrm{CF}}=7.37$, the half-span rule gives $c=8$ with size $0.993=993\alpha$ and the full-span rule
$c=9$ with size $0.904=904\alpha$, and
no support point has upper tail at most $\alpha$ ($c^*=+\infty$, since
$\Pr(d=10)=0.860$). (With the limiting $\gamma_1,\gamma_2$ of the tiers neither shift was
anti-conservative in the sparse uniform cases we computed, $N<40$ with $5\le n\le30$ and
$N\le20$ with $n\le200$; this, too, is only an observation.)

\emph{General null.} Here we recommend the exact law of Section~\ref{sec:algorithm}
whenever it is affordable; for $n=3$ it is an $O(N^2)$ sum (Remark~\ref{rem:n3}). On our
$182$ tables $\Gamma$+CF with the half-span rule had size at most $1.04\alpha$ for $n\ge7$,
and the full-span variant (with the exact moments) stayed below $\alpha$ only for
$n\ge7$ (up to $1.15\alpha$ for $n=5,6$ and $1.30\alpha$ for $n\le4$); but these are $22$
moderately heterogeneous profiles with $\min_iNr_i\ge2.08$, and the dimension $n$ alone
does not control the error. As a starting point for the exact tail check below, the normal
Cornish--Fisher quantile is the better of the two corrected quantiles on these tables at
$1\%$ and $0.1\%$ (Numerical observation~\ref{obs:general}(f)), and $B_3$ is better still at $5\%$ and $1\%$;
none of them is a level guarantee. With small expected counts the corrected quantile can even
leave the support: for $n=7$, $N=21$, $\mathbf q=(1,1,1,10000,1,1,1)$, where six cells
have $Nr_i\approx0.002$, the exact skewness and excess kurtosis are $9.09$ and $83.2$, the
polynomial $C(\cdot;\gamma_1,\gamma_2)$ decreases before $z_{1-\alpha}$, and
$x^{\Gamma+\mathrm{CF}}=-2.18$ and $-10.51$ at $\alpha=1\%$ and $0.1\%$; the rule then
rejects every sample (size $1$), while the exact tests have sizes $0.0084$ and
$7.5\cdot10^{-5}$. For profiles of the same kind with $\min_iNr_i<1$ we found sizes up to
$58\alpha$ with $x$ inside the support (for example $5.8\alpha$ and $58\alpha$ at $1\%$ and
$0.1\%$ for $\mathbf q=(1,1000,1)$, $N=30$).

\emph{The exact tail check.} Under every null, uniform or general, an approximate critical
value should therefore be accepted only if the exact upper tail at the proposed $c$, or a
certified upper bound of it (Proposition~\ref{prop:float}, with correctly rounded weights), is at most $\alpha$; otherwise $c$ is moved up the support until the tail is at most $\alpha$, or
$c^*$ is read off the exact law. Among the deterministic procedures
considered here only this check controls the size, and it is required whenever the size must not exceed $\alpha$. That $z\mapsto C(z;\gamma_1,\gamma_2)$ is
increasing on $[0,z_{1-\alpha}]$ is a cheap diagnostic, but not a sufficient one: it holds
in all $546$ cells of our subgrid and fails in the two size-one cases above and at
$(N,n)=(5,200)$ under the uniform null, but it holds for $\mathbf q=(1,1000,1)$, $N=30$, at
$1\%$, where the size is $5.8\alpha$.

\emph{Cost of the check.} The tail at one point is cheaper than the whole law:
the partial sums
$v=\sum_{i<j}|A(x_{i+1}-x_i)-c_i|$ in~\eqref{eq:dp} do not decrease with $j$, so on the integer axis $V$
of~\eqref{eq:generic} all states with $v\ge C$ can be merged into one, where $C=c$ when the
recurrence runs on the raw statistic $V=d$ and $C=Lc$ when it runs on the centered
$V=L\,d_{\mathbf r}$ (including the uniform null with $V=n\,d$), and the polynomials have
degree $C$ instead of $V_{\max}$; with the running sums one tail then costs $O(nN^2C)$
arithmetic operations and $O(N^2C)$ memory instead of $O(nN^2V_{\max})$ and
$O(N^2V_{\max})$. The
whole procedure avoids the exact law only if the moments and the shift do too: the first
four moments follow from the recurrence~\eqref{eq:dp} with each polynomial in $u$ replaced
by its power sums $\sum_vC_j(t,a,v)\,v^k$, $k\le4$, updated by the binomial theorem
($O(nN^3)$ arithmetic operations; we checked it against the exact law but did not time
it), or by the running sums of Proposition~\ref{prop:runsum} applied to these vectors of
power sums, since $u^s$ acts on them as the binomial map induced by $v\mapsto v+s$
($O(nN^2)$ arithmetic operations), and for $N\ge N_0$ the shift is $s/(2L)$ by Theorem~\ref{thm:lattice}(c); below $N_0$
the shift $s/(2L)$ may be smaller than $H/(2L)$, which the tail check absorbs. In the calibration of this section the shift is half
the actual span $H/L$ of the exact support, not $\frac12$ and not $1/(2L)$;
it equals $s/(2L)$ for $N\ge N_0$ by Theorem~\ref{thm:lattice}(c), and in all $270$ of our
tables, but not for every $N$ (Section~\ref{sec:lattice}).

\section{Open problems}
	\label{sec:open}

\begin{enumerate}
\item \textbf{The step for small $N$.} Theorem~\ref{thm:lattice}(c) gives $s\mid H$ for every
$N$ and $H=s$ for $N\ge N_0$, but equality can fail below $N_0$: for $\mathbf q=(1,8,1)$,
$N=3$, the support of $V$ is $\{12,18,42,72\}$, so $H=6$ while $s=2$, and for
$\mathbf q=(1,3N-1,1)$ one has $H=3\gcd(4,N-1)$ for every $N\ge3$. Characterize the
exceptional pairs $(\mathbf q,N)$ and, for a fixed profile, the least $N_*(\mathbf q)$ with
$H=s$ for all $N\ge N_*(\mathbf q)$; the family above shows that $N_*$ cannot be bounded
independently of $\mathbf q$. In an exhaustive search (all primitive positive profiles with
$n=3$, $L\le16$, $N\le60$; $n=4$, $L\le12$, $N\le40$), every exception with
$N\ge3$ had $N\min_ir_i<\frac13$; whether $N\min_ir_i\ge c$ for an absolute constant $c$
suffices is open.
\item \textbf{Local expansion off the uniform null.} Along a subsequence $N\equiv a\pmod L$,
a local expansion of $\Pr(V=v)$ on the true lattice, uniformly in $v$ with
$|v-\E V|=O(L\sqrt N)$, with a limiting subdensity for each residue class of
$V$ modulo a suitable multiple of $L$, extending Theorem~\ref{thm:n3local}. The classes of Theorem~\ref{thm:residues}
modulo $L$ are not fine enough: for $\mathbf q=(1,2,1)$ and even $N$, $V\equiv0\pmod 4$ always, while the branch
$3|X_2-N/2|$ of Remark~\ref{rem:n3} puts its mass only on multiples of $3$. The multinomial local
expansions~\cite{siotani1984,ouimet2021} are the natural input. For the distribution function the
first lattice correction is covered by Yarnold's expansion for convex sets~\cite{yarnold1972},
since $\{d_{\mathbf r}\le c\}$ is convex; what is open is the push-forward of the local expansions
through the folding to the point probabilities of each residue class, which is what a
per-class continuity correction needs.
\item \textbf{The uniform law for $n=4$ and $n=6$.} The exact uniform law has secondary local
maxima at $n=4$ and $n=6$~\cite[Numerical observation~2.2]{banic_elezovic_p07}, and we have
no mechanism for them. Determine the first non-vanishing periodic term in a local
expansion of the uniform law, separately for each $N$ modulo $n$, and decide whether it
produces the secondary maxima and whether they persist as $N\to\infty$.
\item \textbf{A three-moment reference off the uniform null.} The skewness-fitted beta
improves the calibration under a general null (Numerical observation~\ref{obs:general}(d))
but does not always return
the exact critical value for $n\le5$. Let
$T(c)=\Pr(d_{\mathbf r}\ge c)$, let $c^*$ be the exact critical value and $c^-$ the
preceding support point, so that $T(c^-)>\alpha\ge T(c^*)$. An approximation $\hat T$ of
the upper tail returns $c^*$ as the smallest support point with $\hat T\le\alpha$ whenever
$\max_c|\hat T(c)-T(c)|$, over all support points $c\le c^*$, is
below the decision margin $\min\{T(c^-)-\alpha,\,\alpha-T(c^*)\}$ (when $T(c^*)=\alpha$ the
margin is zero, and only a certified one-sided bound or the exact tail decides); this
margin can be much smaller than half the gap $T(c^-)-T(c^*)$ (for $\mathbf q=(1,2,1)$,
$N=60$, $\alpha=0.1\%$ it is $6.6\cdot10^{-5}$ against $3.4\cdot10^{-4}$). Open is a
three-moment or saddlepoint approximation with a certified one-sided bound on its tail
error, uniformly in a stated range of $N$, $\alpha$ and profiles, so that it either
certifies the decision or tells when the exact tail must be computed. (A size in a fixed
interval such as $[\alpha/2,\alpha]$ cannot be asked of a non-randomized test: in the same
example $T(c^*)=0.39\alpha$; the randomized test attains $\alpha$ exactly.) A candidate is a
saddlepoint approximation~\cite{daniels1954,good1957} based on the cumulant generating
function $N\bigl(K(\mathbf t)-\mathbf t\cdot\Delta\bigr)$ of the centered edge vector
$\tilde D=D-N\Delta$, where
$K(\mathbf t)=\log\bigl(r_1e^{-t_1}+\sum_{j=2}^{n-1}r_je^{t_{j-1}-t_j}+r_ne^{t_{n-1}}\bigr)$
is that of the increment of one draw; the tail of $d_{\mathbf r}$ is then an integral over
polyhedral sign orthants, not a sum of univariate saddlepoint formulas.
\item \textbf{Large $N$ with large $L$.} Exact algorithms whose number of arithmetic
operations does not depend on $L$ exist: enumeration of the $\binom{N+n-1}{n-1}$
compositions, and the $N$-fold convolution of the one-draw law of the edge vector in
Section~\ref{sec:algorithm}, computable by a multidimensional transform of length $2N+1$
per coordinate (exactly by finite-field fast Fourier transforms~\cite{pollard1971} with
Chinese remaindering; characteristic-function and FFT methods for exact multinomial tests
go back to~\cite{baglivo1992,keich2006}). Neither is polynomial jointly in $n$ and $N$: the
transform is exponential in $n$, and the number of compositions, although polynomial in
$n$ for fixed $N$, grows exponentially along joint regimes such as $n\asymp N$. In bit
complexity no exact method can be independent of $L$, since the weights sum to $L^N$. The
open problem is an exact algorithm whose cost is polynomial in
$n$, $N$, $\log L$ and the bit lengths of the rational inputs $c$ and $\alpha$ for
one of two outputs: (i) one tail $\Pr(d_{\mathbf r}\ge c)$, as an exact rational number,
for a given rational $c$; (ii) the exact critical value for a given rational $\alpha$; or a
lower bound showing that none exists. The whole law in output-sensitive form, with cost
polynomial also in the number $m$ of support points rather than in the range
$V_{\max}=O(LN)$ of the recurrence or the factor $\min(L,2^{n-1})$ of the residue
reduction~\cite{banic_elezovic_gct}, is given by the sparse recurrence of
Section~\ref{sec:algorithm} (the proposition on the output-sensitive exact law); it does not
settle (i) and (ii), whose cost is asked to be independent of $m$.
\end{enumerate}

\section*{Conclusion}

We proved the correctness and the cost of the running-sum recurrence for $d$ and
$d_{\mathbf r}$, an error bound for its floating-point version (relative when no
underflow occurs, relative plus absolute otherwise), the lattice of $d_{\mathbf r}$ for
positive rational nulls with $n\ge3$ (exact step $s/L$ for $N\ge N_0$, a containing
lattice for every $N$), the local sawtooth law for the uniform three-cell case and the
limit of the residue-class masses. From the exact tables we read the approach of the cumulants to
their limits and the calibration of continuous approximations. These observations are
limited to positive rational nulls, to the tabulated grids and, for the calibration, to
the exact moments of each table. Our limit results, against which the continuous approximations were judged, hold only
for fixed $n$ and $N\to\infty$ with all $r_i>0$, and even there the fitted laws remain approximations of the limit law;
they give no level guarantee under any null, and they are least reliable when some expected counts are small
or $N$ is small compared with $n$. When the size must not exceed $\alpha$, we
recommend the exact law, or an exact or certified upper tail at the chosen critical value.

\medskip\noindent\textbf{Acknowledgements.} The observation behind the proposition on the
output-sensitive exact law (Section~\ref{sec:algorithm}) is due to an external review of an
earlier version of this paper. The authors used AI assistants in the preparation of this
paper: Anthropic's Claude Opus~5.5 for additional literature searches, symbolic and numerical
verification, and editorial support, and OpenAI's Codex Sol~6.1 for a critical review of an
earlier version and for assistance with the subsequent revision. All mathematical results
and all changes to the text were checked independently by the authors, who take full
responsibility for the content.

\medskip\noindent\textbf{Code and data.} The scripts in exact integer arithmetic
(brute-force enumeration, the recurrence~\eqref{eq:dp} and the running sums) produce every
exact value in the paper; separate scripts run the floating-point version, the quadratures
and the Genz cross-check.
The programs and exact tables used in this paper are available from the authors on
request, and a common archive for the series, with the companion paper on tables and
software~\cite{banic_elezovic_p11}, is being prepared.

\end{document}